\documentclass[11pt,fleqn]{article}

\usepackage{a4}
\usepackage{amstext}
\usepackage{amsfonts}
\usepackage{amssymb}
\usepackage{color}
\usepackage{epsfig}
\usepackage{verbatim}
\usepackage[dvips]{hyperref}
\graphicspath{{./Figs/}}

\def\d{d}
\def\f{I}
 
\def\om{\omega}

\begin{document}

\begin{flushright}
HU-EP-11/50 \\
SFB/CPP-11-59 \\
DESY 11-211
\end{flushright}

\begin{center}

{\huge \bf Confining dyon gas \\
\vspace{0.2cm}
with finite-volume effects under control}

\vspace{0.5cm}

\textbf{Falk Bruckmann$^{\dagger}$, Simon Dinter$^{\ddag *}$, 
        Ernst-Michael Ilgenfritz$^{*\sharp}$, Benjamin Maier$^{*}$, 
        Michael M\"uller-Preussker$^{*}$, Marc Wagner$^{*\triangle}$} \\~\\

${}^{\dagger}$ Universit\"at Regensburg, Institut f\"ur Theoretische Physik, 
       D-93040 Regensburg, Germany

${}^{\ddag}$ NIC, DESY Zeuthen, Platanenallee 6, D-15738 Zeuthen, Germany

${}^{*}$ Humboldt-Universit\"at zu Berlin, Institut f\"ur Physik, \\
        Newtonstr. 15, D-12489 Berlin, Germany

${}^{\sharp}$ Joint Institute for Nuclear Research, VBLHEP, 
        141980 Dubna, Russia

${}^{\triangle}$ Goethe-Universit\"at Frankfurt am Main, 
        Institut f\"ur Theoretische Physik, \\ 
        Max-von-Laue-Stra{\ss}e 1, D-60438 Frankfurt am Main, Germany \\~\\

\texttt{falk.bruckmann@physik.uni-regensburg.de} \\
\texttt{dinter@physik.hu-berlin.de} \\
\texttt{ilgenfri@physik.hu-berlin.de} \\
\texttt{bfmaier@physik.hu-berlin.de} \\
\texttt{mmp@physik.hu-berlin.de} \\
\texttt{mwagner@th.physik.uni-frankfurt.de}

\vspace{0.5cm}

November 10, 2011

\end{center}

\vspace{0.1cm}

\begin{tabular*}{16cm}{l@{\extracolsep{\fill}}r} \hline \end{tabular*}

\vspace{-0.4cm}
\begin{center} \textbf{Abstract} \end{center}
\vspace{-0.4cm}

As an approach to describe the long-range properties of non-Abelian 
gauge theories at non-zero temperature $T < T_c$, we consider a non-interacting 
ensemble of dyons (magnetic monopoles) with non-trivial holonomy. We show 
analytically, that the quark-antiquark free energy from the Polyakov loop 
correlator grows linearly with the distance, and how the string tension scales with the dyon density. In numerical treatments, the long-range 
tails of the dyon fields cause severe finite-volume effects. Therefore, we 
demonstrate the application of Ewald's summation method to this system. 
Finite-volume effects are shown to be under control, which is a crucial 
requirement for numerical studies of interacting dyon ensembles.

\begin{tabular*}{16cm}{l@{\extracolsep{\fill}}r} \hline \end{tabular*}

\thispagestyle{empty}

\newpage

\setcounter{page}{1}

\section{Introduction}
\label{sec:introduction}

Insight into the mechanisms of the QCD vacuum is not only provided by 
simulations of lattice gauge theory -- an ab initio method, whose numerical 
results, however, are hard to interpret -- but also by analytical 
non-perturbative approaches like the semiclassical one~\cite{Callan:1977gz,
Callan:1978bm}. The latter relies for instance on instantons, selfdual and 
anti-selfdual solutions of the Euclidean Yang-Mills 
equations~\cite{Belavin:1975fg}. Instantons 
in $\mathbb{R}^4$ are localized in space \emph{and time}, but also naturally 
contain long-range fields (since the Yang-Mills Lagrangian is scale-invariant): 
the gauge potential $A_\mu$ decays like the inverse of the four-dimensional 
distance to their center or its third power, in the regular and singular gauges, 
respectively. Semiclassically motivated models of the QCD vacuum based on 
instantons are suitable to describe certain non-perturbative effects like 
chiral symmetry breaking, but so far cannot explain confinement. For more 
details of instanton models we refer to the 
reviews~\cite{Schafer:1996wv,Diakonov:2002fq}.  

When studying instantons or similar long-range (or ``infrared'') objects 
in a finite-volume approximation -- an unavoidable restriction for virtually 
every numerical approach -- one expects severe effects: interactions with 
objects outside the finite volume (and their contribution to observables) 
are neglected, which can introduce considerable systematic deviations from 
analogous systems with infinite extent.

The purpose of our work is two-fold. On the one hand, we investigate 
\emph{confinement in a semiclassical approach at non-zero temperature}.
Guided by the invention of KvBLL-calorons with non-trivial 
holonomy~\cite{Kraan:1998pm,Kraan:1998sn,Lee:1998bb}
our basic objects are dyons - the constituents of calorons. We assume 
maximally non-trivial holonomy in order to describe the confinement phase of 
the model. Dyons will be analytically shown to provide a confining Polyakov 
loop correlator already within the simplest non-interacting model for the 
low-temperature phase.

Concerning the long-range nature, dyons are as difficult to simulate as 
instantons. Therefore, as the second part, we provide the proof-of-concept 
for \emph{a method capable to control the finite-volume effects} in such 
systems in an efficient way: Ewald's summation method~\cite{Ewald:1921}. 
This method was originally developed for Coulomb interactions typical, 
for example, in plasma or soft matter physics.
When using Ewald's method, the infinite space is mimicked by infinitely 
many replicas of one so-called ``supercell'' that contains a (for numerical 
simulations) feasible number of objects/charges. Typical observables like 
potentials are sums, which can be split into a short-range and a long-range 
part. After rewriting the long-range sum by means of a Fourier transform, 
both sums can efficiently be computed (see Section~\ref{subsec:outline} for 
details). In order to come  back to the original system in the infinite space, 
only the volume of the supercell has to be extrapolated to infinity at the end
of a computation. In Ewald's method this is a well-controlled limit in 
contrast to simpler approaches. 

In this work we apply this method to the simplest dyon model, for which a 
comparison with analytical results can be made. The advantages of the 
Ewald method will be essential for later numerical studies of interacting dyon 
ensembles.

The paper is organized as follows. In Section~\ref{sec:dyons} the crucial 
features of dyons are introduced. The Polyakov loop correlator in a 
non-interacting dyon model is analytically evaluated in 
Section~\ref{sec:analytical}, both in infinite and finite volume. Contact 
to lattice simulations is made and consequences for dyon models are discussed.
Section~\ref{sec:ewald} introduces Ewald's summation method.
In Section~\ref{sec:results} numerical results are presented. 
Section~\ref{sec:summary} summarizes this work and opens a view to 
simulations of interacting dyon models. 
In Appendix~\ref{app:integrals} some integrals required in the 
analytical approach are computed, whereas Appendix~\ref{subsec:ewald_vs_other} 
compares Ewald's method of summing over an infinite number of copies of 
the supercell with the result of the converging series of sums over a 
finite number of cells.

\section{Dyon gas model for $SU(2)$ Yang-Mills theory
\label{sec:dyons}}

The notion of finite temperature instanton solutions, traditionally called 
calorons~\cite{Harrington:1978ve}, has been radically extended when new 
caloron solutions were found by Kraan and van 
Baal~\cite{Kraan:1998pm,Kraan:1998sn} as well as Lee and Lu~\cite{Lee:1998bb}.
They consist of magnetic monopoles as constituents. The latter also carry 
(the Euclidean analog of) electric charge and will therefore be called dyons. 
The asymptotic Polyakov loop of these solutions, the trace of the 
so-called holonomy, is an additional external parameter that governs for 
instance how the instanton (caloron) action is shared by the constituent 
dyons.

Dyons as selfdual objects at finite temperature can be obtained by
considering the gauge field of a caloron in the limit of infinite dyon 
separation \cite{Kraan:1998pm}. The dyon constituents can be understood 
as BPS monopoles interpreting the scalar Higgs field as a
temporal gauge field. In the far-field limit, when the distance to the
dyon center is large, the gauge field is Abelian along the direction 
of the asymptotic Polyakov loop (``the color direction of the Higgs field''), 
which we take diagonal
\begin{eqnarray}
& & A_0 \ \ \to \ \ 2\pi\om T \ \sigma_3 , \label{eqn:asympt_Polloop_def_upper} \\
& & P(\mathbf{r}) \ \equiv 
\ \frac{1}{2} \textrm{Tr}\bigg(\exp\bigg(i\int_0^{1/T} dx_0 \, 
         A_0(x_0,\mathbf{r})\bigg)\bigg) \ \ \to \ \
\frac{1}{2}\textrm{Tr}\Big(\exp\Big(2\pi i \om\sigma_3\Big)\Big) \ = 
\ \cos(2\pi\om) ,
\label{eqn:asympt_Polloop_def}
\end{eqnarray}
with $T$ the temperature and $\sigma_3=\mbox{diag}(+1,-1)$ the third Pauli 
matrix. The parameter $\om$ specifies the holonomy. Maximally non-trivial 
holonomy refers to $\om=1/4$ and $P(\mathbf{r})\to 0$ and is conjectured to 
be valid in the confined phase, where $\langle P \rangle = 0$ (as a quantum 
spatial average), in contrast to trivial holonomy $P(\mathbf{r})\to \pm 1$
valid deep in the deconfined phase. The viability of confinement has been 
shown semi-analytically, even without complete decomposition into constituents 
\cite{Gerhold:2006sk}. Further investigations supporting the conjecture have 
been focused on the quantum amplitude~\cite{Diakonov:2004jn}, moduli space 
metric~\cite{Kraan:1998pn,Diakonov:2005qa} and the vortex content of 
calorons~\cite{Bruckmann:2009pa}.

Dyons are genuine non-Abelian objects, whose field components 
color-per\-pen\-di\-cu\-lar to the asymptotic Polyakov loop
decay exponentially (like e.g.\ fields of massive bosons color perpendicular 
to the Higgs vacuum expectation value) outside a region of size $\beta \equiv 1/T$. 
The dyons' long-range fields are Abelian in the same color direction and 
Coulomb-like (in addition to the constant of 
Eq.~(\ref{eqn:asympt_Polloop_def_upper})):
\begin{eqnarray}
\label{eqn:EQN010}
a_0(\mathbf{r};q) \ \ = \ \ \frac{q}{r} \: , \quad
a_1(\mathbf{r};q) \ \ = \ \ -\frac{q y}{r (r-z)} \: , \quad
a_2(\mathbf{r};q) \ \ = \ \ +\frac{q x}{r (r-z)} \: , \quad
a_3(\mathbf{r};q) \ \ = \ \ 0 \:,
\end{eqnarray}
where $\mathbf{r} = (x , y , z)$ and $r = |\mathbf{r}|$ is the
three-dimensional distance to the dyon center.
With the help of 't Hooft's symbol one can write in a compact way
\begin{eqnarray}
a_\mu(\mathbf{r};q) \ \ = \ \ -q \bar{\eta}_{\mu \nu}^3 \partial_\nu \ln (r-z) .
\end{eqnarray}
The vector potential in this limit results in electric and magnetic fields
\begin{eqnarray}
\mathbf{e} \ \ = \ \ \frac{q \mathbf{r}}{r^3} \quad , 
\quad \mathbf{b} \ \ = \ \ q \,\bigg(\frac{\mathbf{r}}{r^3} + 
 4 \pi \delta(x) \delta(y) \Theta(z) \mathbf{e}_z\bigg) . 
\end{eqnarray}
The possible charges are $q=+1$ for dyons and $q = -1$ for anti-dyons. 
The Dirac string singularities along the positive $z$-axis are artefacts
of the Abelian limit. They do not need to concern us here.

So far we have considered selfdual dyons, whose electric and magnetic
charges are coupled as $\mathbf{e} = \mathbf{b}$ (neglecting Dirac strings).
The actual semiclassical field content dominating the partition function 
should be built 
from selfdual and antiselfdual dyons and antidyons. For antiselfdual 
dyons and antidyons, for which $\mathbf{e} = - \mathbf{b}$, the 't Hooft 
symbols $\bar{\eta}_{\mu \nu}^3$ are replaced by $\eta_{\mu \nu}^3$.
As we will argue below, all that matters for our work is the Coulomb-like
decay of $a_0$ away from positive and negative electric charges $\pm q$, which 
are placed at random positions. 
This will apply also to a mixed model including antiselfdual dyons and 
antidyons. In other words, for the aspects under study selfduality and  
antiselfduality and the magnetic charges are irrelevant properties and as a consequence, our 
formulation respects CP-invariance.
In due course, total numbers and densities will refer to dyons of both magnetic
charges.

The superposition of the gauge fields of $2 K$ dyons in the Abelian 
limit reads
\begin{eqnarray}
A_\mu(\mathbf{r}) \ \ = \ \ \bigg(\delta_{\mu 0} 2\pi\om T
+ \frac{1}{2} \sum_{i=1}^K \sum_{m=1}^2 a_\mu(\mathbf{r} -
\mathbf{r}_i^m;q_m)\bigg)\sigma_3 ,
\label{eqn:Amu}
\end{eqnarray}
where $\mathbf{r}_i^m$ and $q_m=-(-1)^m$ are the positions and charges
of the $i$-th dyon ($m=1$) and antidyon ($m=2$), respectively.

Like the vector potentials, interactions of monopoles or dyons behave 
Coulomb-like~\cite{Manton:1985hs,Gibbons:1986df,Gibbons:1995yw}.
Relying on these long-range fields, Diakonov and Petrov have presented a 
formal solution for the statistical mechanics of purely (anti)selfdual
dyons~\cite{Diakonov:2007nv}, later extended to both 
selfdualities~\cite{Diakonov:2009jq}. 

The assumed moduli space metric of the dyon configurations allowed for 
a particular analytic treatment in the spirit of Polyakov's monopole 
confinement mechanism~\cite{Polyakov:1976fu}.
In an attempt to implement a simulation for dyon gases with this 
interaction, however, we have noticed that the metric severely suffers 
from non-positivity~\cite{Bruckmann:2009nw}, which casts 
doubts on the validity of the analytical results obtained 
in~\cite{Diakonov:2007nv} in the context of Yang-Mills theory.

In this paper we consider dyon ensembles without moduli space metric 
or other interactions, i.e.\ we perform a uniform sampling of dyon 
positions. We will focus on maximally non-trivial holonomy, $\om=1/4$, 
where both dyons and antidyons possess the same topological
charge of $2\om=1-2\om=1/2$ of an instanton unit and hence the same 
action, such that they do not differ in their classical and quantum weight.
Therefore, it is natural to use the same number of dyons and antidyons,
i.e.\ an electrically and magnetically neutral ensemble and denote by
$n_D=2K$ the total number of dyons and antidyons.  
For other values of the holonomy, say for those close to maximally 
non-trivial, the assumption of equally frequent dyons is only a first 
approximation, arguments suggesting the contrary are discussed in 
references~\cite{Bornyakov:2008im,Bruckmann:2009ne}.

The basic parameters of our model are the holonomy $\om$, the 3-dimensional 
density of dyons $\rho$ and the temperature $T$. The scale can be set by 
identifying the string tension $\sigma$ extracted from the free energy of 
a static quark-antiquark pair with the corresponding lattice result 
as explained in Section~\ref{subsec:calibration}. 

Our primary observable is the local Polyakov loop $P(\mathbf{r})$
at position $\mathbf{r}$ (cf.\ Eq.~(\ref{eqn:asympt_Polloop_def})).
In the Abelian limit the fields are static and we need to sum the holonomy 
and the $a_0$-component of the individual dyons as follows,
\begin{eqnarray}
P(\mathbf{r}) \ \ = \ \ \cos \bigg( 2\pi\om+\frac{1}{2T} \Phi(\mathbf{r})\bigg)
\: , \qquad
P(\mathbf{r})\bigg|_{\om=1/4} \ \ = 
\ \ -\sin \bigg(\frac{1}{2T} \Phi(\mathbf{r})\bigg)
\label{eqn:P}
\end{eqnarray}
with the following sum over Coulomb terms
\begin{eqnarray}
\Phi(\mathbf{r}) \ \ \equiv \ \  \sum_{i=1}^K \sum_{m=1}^2
\frac{q_m}{|\mathbf{r} - \mathbf{r}_i^m|} 
=\sum_{i=1}^K \bigg[
\frac{1}{|\mathbf{r} - \mathbf{r}_i^1|}-
\frac{1}{|\mathbf{r} - \mathbf{r}_i^2|} \bigg] \; .
\label{eqn:Phi}
\end{eqnarray}
As well-known, the correlator of Polyakov loops yields the free energy of a 
static quark-antiquark pair:
\begin{eqnarray}
F_{\bar{Q} Q}(\d) \ \ = \ \ -T \ln \Big\langle P(\mathbf{r})
P^\dagger(\mathbf{r'}) \Big\rangle \quad  ,
\quad \d \ \ \equiv \ \ |\mathbf{r}-\mathbf{r'}| \; .
\label{eqn:correlator}
\end{eqnarray}
From the point of view of a Coulomb gas, correlators of trigonometric 
functions are slightly exotic, but for the dyon model of QCD this is the 
essential correlation function probing confinement. 

In simulations using a finite number of dyons and anti-dyons the positions 
$\mathbf{r}_i^m$ are restricted to a finite dyon sampling volume. Then contributions 
from dyons outside this volume to the sum in Eq.~(\ref{eqn:Phi}) are 
ignored. How one can control such 
finite-volume effects systematically, is the main subject of the second part 
of this paper. We will resort to Ewald's summation method and compare it to 
the analytic result for Polyakov loop correlators, which are presented in 
the next section.

\section{The Polyakov loop correlator in a non-interacting dyon gas model} 
\label{sec:analytical}

In this section we treat the non-interacting dyon ensemble analytically. 
In particular we show the Polyakov loop correlator (\ref{eqn:correlator}) 
from random dyons to be confining and investigate finite-volume effects. 
Interacting dyon ensembles can be reformulated as scalar theories 
\cite{Polyakov:1978vu,Diakonov:2007nv}, but here -- due to the absence of 
interactions -- the model can be solved. In this simple system we 
therefore obtain analytic formulae for the string tension, which later 
will be used to set the scale and as a benchmark for numerical methods.

\subsection{The correlator}
\label{sec:correlator}

Expectation values of observables  $O$ in the ensemble with $K$ dyons 
of charge $+1$ at positions $\mathbf{r}_i^{1}$ and $K$ dyons of charge 
$-1$ at positions $\mathbf{r}_i^{2}$ are given by:
\begin{eqnarray}
 \Big\langle O \Big\rangle \ \  = \ \ 
\int\prod_{i=1}^K d\mathbf{r}_i^1d\mathbf{r}_i^2 \,O\Big(\{\mathbf{r}_i^1,
              \mathbf{r}_i^2\}\Big)\Big/
\int\prod_{i=1}^K d\mathbf{r}_i^1d\mathbf{r}_i^2=
\int\prod_{i=1}^K d\mathbf{r}_i^1d\mathbf{r}_i^2 \,O\Big(\{\mathbf{r}_i^1,
              \mathbf{r}_i^2\}\Big)\Big/V^{2K}
\end{eqnarray}
where $V$ is the spatial volume in which the $2K$ dyons are randomly 
distributed. Their density is 
\begin{eqnarray}
 \rho \ \ = \ \ \frac{2K}{V}
\end{eqnarray}
accordingly.

The Polyakov loop correlator is given by a product of cosines, see (\ref{eqn:P}), 
(\ref{eqn:Phi}) and (\ref{eqn:correlator}), and can be rewritten as
\begin{eqnarray}
\Big\langle P(\mathbf{r}) P(\mathbf{r'}) \Big\rangle \ \ = \ \ 
\frac{1}{2}\,
\Big\langle \cos \Big(4\pi\om+\frac{\Phi_+}{2T}\Big)\Big\rangle
+\frac{1}{2}\,\Big\langle\cos \Big(\frac{\Phi_-}{2T}\Big) \Big\rangle\,,
\label{eqn:correlator_first}
\end{eqnarray}
where the Coulomb sums
\begin{eqnarray}
\Phi_\pm \ \ \equiv \ \ 
\Phi(\mathbf{r})\pm \Phi(\mathbf{r'})
\ \ = \ \ 
\sum_{i=1}^K \bigg[\Big(
\frac{1}{|\mathbf{r} - \mathbf{r}_i^1|}\pm 
\frac{1}{|\mathbf{r'} - \mathbf{r}_i^1|} \Big)-
\Big(
\frac{1}{|\mathbf{r} - \mathbf{r}_i^2|}\pm 
\frac{1}{|\mathbf{r'} - \mathbf{r}_i^2|} \Big)\bigg]
\end{eqnarray}
contain all dyons and depend on the two measurement points 
$\mathbf{r}$ and $\mathbf{r'}$.

Rewriting 
\begin{eqnarray}
 \Big\langle P(\mathbf{r}) P(\mathbf{r'}) \Big\rangle \ \ = \ \ 
\frac{1}{4}e^{4\pi i\om}\,\Big\langle \exp\Big(i\frac{\Phi_+}{2T}\Big)\Big\rangle
+\mbox{c.c.}+
\frac{1}{4}\Big\langle \exp\Big(i\frac{\Phi_-}{2T}\Big)\Big\rangle+\mbox{c.c.}
\end{eqnarray}
the ingredients are the following expectation values
\begin{eqnarray}
 \Big\langle \exp\Big(i\frac{\Phi_\pm}{2T}\Big)\Big\rangle \ \ & = & \ \  
\frac{1}{V^{K}}
\int\prod_{i=1}^Kd\mathbf{r}_i^1\exp\bigg[\frac{i}{2T}
\Big(\frac{1}{|\mathbf{r}-\mathbf{r}_i^1|}\pm\frac{1}{|\mathbf{r'}-
 \mathbf{r}_i^1|}\Big)\bigg]\times \mbox{c.c.}\nonumber\\
\ \ & = & \ \  
\bigg(\frac{1}{V}
\int \!d\mathbf{s}\,\exp\bigg[\frac{i}{2T}
\Big(\frac{1}{|\mathbf{r}-\mathbf{s}|}\pm\frac{1}{|\mathbf{r'}-
\mathbf{s}|}\Big)\bigg]\bigg)^K\times \mbox{c.c.}
\ \ = \ \ \left(\frac{|\f_{\pm}|}{V}\right)^{2K}\,.
\end{eqnarray}
They are real and have factorized into integrals given in terms of one 
dyon location only:
\begin{eqnarray}
 \f_\pm \ \ \equiv \ \ \int \!d\mathbf{s}\,\exp\bigg[\frac{i}{2T}
\Big(\frac{1}{|\mathbf{r}-\mathbf{s}|}\pm\frac{1}{|\mathbf{r'}-
\mathbf{s}|}\Big)\bigg] .
\label{eqn:fpm}
\end{eqnarray}
The result for the Polyakov loop correlator is then
\begin{eqnarray}
\Big\langle P(\mathbf{r}) P(\mathbf{r'}) \Big\rangle  
\ \ &  = & \ \ 
\frac{1}{2}\,\cos(4\pi\om)
\left(\frac{|\f_+|}{V}\right)^{2K}
+\frac{1}{2}\left(\frac{|\f_-|}{V}\right)^{2K}\,.
\label{eqn:correlator_second}
\end{eqnarray}
Keeping the density fixed, we can replace the number of 
dyons $2K$ and obtain
\begin{eqnarray}
\Big\langle P(\mathbf{r}) P(\mathbf{r'}) \Big\rangle  
\ \ &  = & \ \  
\frac{1}{2}\,\cos(4\pi\om)
\bigg[\bigg(1+\frac{|\f_+|-V}{V}\bigg)^{V}\bigg]^\rho
+\frac{1}{2}
\bigg[\bigg(1+\frac{|\f_-|-V}{V}\bigg)^{V}\bigg]^\rho\,,
\label{eqn:correlator_third_gen}
\end{eqnarray}
in particular at maximally non-trivial holonomy $\om=1/4$
\begin{eqnarray}
\Big\langle P(\mathbf{r}) P(\mathbf{r'}) \Big\rangle\bigg|_{\om=1/4}  
\ \ &  = & \ \  
-\frac{1}{2}
\bigg[\bigg(1+\frac{|\f_+|-V}{V}\bigg)^{V}\bigg]^\rho
+\frac{1}{2}
\bigg[\bigg(1+\frac{|\f_-|-V}{V}\bigg)^{V}\bigg]^\rho\,.
\label{eqn:correlator_third}
\end{eqnarray}
This (still exact) form with the explicit volume 
dependence\footnote{The use of dimensionful quantities in the exponent 
can be avoided by normalizing with some standard volume.} 
has been chosen in anticipation of the properties of $\f_\pm$ 
discussed below. 

\subsection{String tension in the infinite-volume limit}
\label{subsec:confinf_volume}

The task here will be to calculate the asymptotic behavior of the 
integrals $\f_\pm$ of Eq.\ (\ref{eqn:fpm}) in the limit of 
large quark-antiquark separations. The behavior at finite separations 
as well as finite-volume corrections are investigated in the next subsection.

By shifting and rotating the integration variable in Eq. (\ref{eqn:fpm})
one can see that $\f_\pm$ are functions of the distance 
$ |\mathbf{r}-\mathbf{r'}| = d $ as expected.

The integrands of both integrals $\f_+$ and $\f_-$ asymptotically approach 
unity, the corresponding (divergent) term will be canceled by $V$ in 
Eq. (\ref{eqn:correlator_third}).
However, there is an important difference: with the relative plus sign 
in $\f_+$ the next term in the asymptotic expansion is the monopole term 
(proportional to $2/s$), while the integrand of $\f_-$ will only start 
with a dipole term due to the relative minus sign. We will show that 
as a consequence the first term in Eq. (\ref{eqn:correlator_third}) 
vanishes in the infinite-volume limit, whereas the second term survives 
and induces the string tension.

We consider regularized integrals in a 3-ball
of radius $R$ and fix the Polyakov loop arguments at $\mathbf{r}=(0,0,+d/2)$ 
and $\mathbf{r'}=(0,0,-d/2)$. Notice first that the integration variable 
$\mathbf{s}$ can be rescaled by the temperature
\begin{eqnarray}
I_\pm \ \ = \ \ \frac{1}{T^3} \underbrace{\int_{S^3_{R T}} d\mathbf{s} \, 
\exp\bigg(\frac{i}{2} \bigg(\frac{1}{|\mathbf{r} T-\mathbf{s}|} \pm 
\frac{1}{|\mathbf{r}' T-\mathbf{s}|}\bigg)\bigg)}_{= f_\pm}
\label{eqn:fpm_rescaled}
\end{eqnarray}
such that these integrals are functions of the finite-volume radius $R$ 
and the separation $d$ only, both in units of $1/T$.
In other words
\begin{eqnarray}
I_\pm \ \ = \ \ \frac{1}{T^3} f_\pm(d T,R T)\,.
\label{eqn:form_Is}
\end{eqnarray} 
In spherical coordinates the $dT$ dependence of the distances
$|\mathbf{r} T-\mathbf{s}|$ and $|\mathbf{r}' T-\mathbf{s}|$ becomes explicit:
\begin{eqnarray}
D_{\pm}(s,\theta,dT) \equiv |(0,0,\pm d/2)T-\mathbf{s}|=
       \sqrt{s^2 \mp s d T\cos\theta + (d T)^2/4} \,.
\label{eqn:abs_values}
\end{eqnarray} 

In order to evaluate the leading terms in the radius $R$, we consider the 
$RT$-derivatives of $f_\pm$ for large $RT$ given by angle integrals on the 
2-sphere $s=RT$:
\begin{eqnarray}
 \frac{d}{d(RT)}f_\pm & =&  2\pi (RT)^2
\int_0^\pi d \theta \, \sin\theta\\
&&\exp\bigg(\frac{i}{2} 
\bigg(\frac{1}{RT\sqrt{1+d/R\cdot\cos\theta+d^2/4R^2}}
\pm
\frac{1}{RT\sqrt{1-d/R\cdot\cos\theta+d^2/4R^2}}~\bigg)\bigg) .\nonumber
\end{eqnarray}
This can be expanded in $1/RT$ and $d/R$ to give
\begin{eqnarray}
 \frac{d}{d(RT)}f_+ & = & 
4\pi(RT)^2+i4\pi (RT)-2\pi+O\left(\frac{1}{RT},\frac{d^4}{R^4}\right)\\
 \frac{d}{d(RT)}f_- & = & 4\pi(RT)^2+
O\left(\frac{d^2}{R^2},\frac{1}{(RT)^2}\frac{d^4}{R^4}\right) ,
\label{eqn:result_fs}
\end{eqnarray}
where the first terms on the right hand sides will, of course, be the 
volume contributions. Moreover, the aforementioned difference in the two 
integrals concerning subleading terms is clearly visible.
Integrating back with respect to $RT$ then yields
\begin{eqnarray}
 f_+ & = & \frac{4\pi}{3}(RT)^3+i2 \pi(RT)^2-2\pi (RT)+O(\ln RT)+g_+(dT) 
\label{eqn:Rdep_plus}\\
 f_- & = & \frac{4\pi}{3}(RT)^3+g_-(dT)\, 
\label{eqn:Rdep_minus}
\end{eqnarray}
where $g_\pm$ are $RT$-independent and where we have neglected all terms 
vanishing as $RT\to\infty$. 

For $|\f_+|$ we finally get the following leading 
terms:
\begin{eqnarray}
 |\f_+| \ \  =  \ \ \left|\frac{4\pi}{3} R^3-2\pi \frac{R}{T^2} + 
i\cdot 2 \pi \frac{R^2}{T}\right|
\ \  =  \ \ V-cT^2R\,,\qquad
 \frac{|\f_+|-V}{V} \ \  =  \ \ -c'\frac{T^2}{R^2}
\end{eqnarray}
with $c$ and $c'$ being positive constants.
In the infinite-volume limit at fixed temperature the contribution to 
the Polyakov loop correlator vanishes 
\begin{eqnarray}
 \lim_{V\to\infty}\left(1+\frac{|\f_+|-V}{V}\right)^{V}=
\lim_{R\to\infty}\left(1-\frac{c'T^2}{R^2}\right)^{\frac{4\pi}{3} R^3} \ \ 
 = \ \ 0 .
\label{eqn:result_plus}
\end{eqnarray}
In $|\f_-|$, on the other hand, only $R$-independent terms enter the Polyakov loop 
correlator as
\begin{eqnarray}
 |\f_-| \ \ & = & \left|V+\frac{g_-}{T^3}\right| = V+\frac{g_-}{T^3} 
\end{eqnarray}
and 
\begin{eqnarray}
 \lim_{V\to\infty}\left(1+\frac{|\f_-|-V}{V}\right)^{V}=
 \lim_{V\to\infty}\left(1+\frac{g_-/T^3}{V}\right)^{V}=
\exp\left(\frac{g_-}{T^3}\right) .
 \label{eqn:result_minus}
\end{eqnarray}
Hence it remains to compute $g_-$ as a function of the Polyakov loop 
separation $d=|\mathbf{r}-\mathbf{r'}|$ 
(in units of $T$), which according to the above is
\begin{eqnarray}
 g_-=\int_{\mathbb{R}^3} d\mathbf{s} \, \bigg\{\exp\bigg(\frac{i}{2} 
\bigg(\frac{1}{|\mathbf{r} T-\mathbf{s}|} - 
\frac{1}{|\mathbf{r'} T-\mathbf{s}|}\bigg)\bigg)-1\bigg\} .
\end{eqnarray}
The imaginary part vanishes by invariance under reflections  
$\mathbf{s}\to-\mathbf{s}$. We split
\begin{eqnarray}
 g_-  =  -\frac{1}{8}\,g_-^{(2)}+g_-^{({\rm res})}\,, \qquad
 g_-^{(2)}  \equiv  \int_{\mathbb{R}^3} d\mathbf{s} 
 \bigg(\frac{1}{|\mathbf{r} T-\mathbf{s}|} - 
\frac{1}{|\mathbf{r'} T-\mathbf{s}|}\bigg)^2
\label{eqn:splitting}
\end{eqnarray}
such that all terms are integrable around $\mathbf{r}T$ and $\mathbf{r'}T$.
The important observation is now that the second-order contribution is 
linear in $dT$ (see also \cite{Diakonov:2004jn}),
\begin{eqnarray}
 g_-^{(2)} = 4\pi dT\qquad 
\label{eqn:result_splitting}
\end{eqnarray}
whereas the remainder $g_-^{({\rm res})}$ is bound by a constant independent 
of $dT$, both derived in detail in Appendix~\ref{app:integrals}.

Finally, in the Polyakov loop correlator (\ref{eqn:correlator_third}), using 
Eqs. (\ref{eqn:result_plus}), (\ref{eqn:result_minus}), (\ref{eqn:splitting}), and 
(\ref{eqn:result_splitting}), we obtain an exponential decay 
at large distance $d=|\mathbf{r}-\mathbf{r'}|$
\begin{eqnarray}
 \Big\langle P(\mathbf{r}) P(\mathbf{r'}) \Big\rangle  
\ \  =  \ \ 
\frac{1}{2}\,
\exp\Big(-\frac{\pi d \rho }{2T^2}+\mbox{const.}\Big)
\end{eqnarray}
or equivalently a linear growth of the free energy
\begin{eqnarray}
 F_{\bar{Q} Q}(\d)\ \   =  \ \ \sigma\, d + \mbox{const.}
 \label{eqn:free_infvolume}
\end{eqnarray}
and read off the string tension
\begin{eqnarray}
 \sigma \ \ = \ \ \frac{\pi}{2}\,\frac{\rho}{T}\,.
 \label{eqn:string_infvolume}
\end{eqnarray}
Given the dependence on $\rho$, $d$ and $T$ in Eqs. (\ref{eqn:correlator_third}), 
(\ref{eqn:form_Is}), (\ref{eqn:result_fs}), and (\ref{eqn:result_minus}), the 
coefficient of a term linear in $d$ can only be of that form (also for dimensional 
reasons). The achievement of this part of our work was to analytically prove this 
confining behavior and to determine the proportionality factor. 

As a side result we find that the holonomy dependence has dropped out completely
in the infinite volume (technically because $\om$ enters together with $\f_+$, 
see Eq. (\ref{eqn:correlator_third_gen}), this contribution, however, vanishes 
in the infinite-volume limit). This is consistent with the fact that the average 
Polyakov loop in our model actually vanishes for all holonomies in the 
infinite-volume limit, which is not difficult to show. 

In other words, the disorder generated by long-range fields of dyons dominates 
the effect of the holonomy on the average Polyakov loop. We remind the reader that 
this finding is based on the same density of all kinds of dyons for all 
holonomies. Hence our model is valid only at maximally non-trivial holonomy, 
i.e.\ in the low temperature phase, whereas in the high-temperature phase 
modifications are expected that may reintroduce a holonomy dependence.
 
\subsection{Fixing the physical scale}  
\label{subsec:calibration}

With the string tension of Eq.~(\ref{eqn:string_infvolume}) at hand, we can 
set the scale of our model. All analytical and later numerical calculations 
provide, of course, relations between dimensionless quantities. As we have 
done already, we can measure all lengths in units of the inverse temperature 
$\beta = 1/T$. 
For the string tension this means
 \begin{eqnarray}
  \frac{\sigma}{T^2} \ \ = \ \ \frac{\pi}{2}\,\frac{\rho}{T^3}
   \ \ = \ \ \frac{\pi}{2}\,\left(\frac{\beta}{\rho^{-1/3}}\right)^3
  \ \ \equiv \ \ \frac{\pi}{2} (f_P)^3 \,.
\label{eqn packing}
 \end{eqnarray}
The ratio on the left hand side is known from lattice simulations.
We have introduced a ``packing fraction'' $f_P$ of the dyon gas, since 
$\rho^{-1/3}$ represents the mean distance and $\beta$ can be interpreted 
as being proportional to the core-size of corresponding non-Abelian dyons.

We resort to lattice results on the $SU(2)$ string tension and its 
temperature dependence in \cite{Digal:2003jc}. We parameterize these results 
(cf.\ Fig.~3 in that reference) by
\begin{eqnarray}
\label{EQN666} 
\frac{\sigma(T)}{\sigma(T=0)} \ \ = 
\ \ A \bigg(1-\frac{T}{T_c}\bigg)^{0.63} \bigg(1 + 
            B \bigg(1-\frac{T}{T_c}\bigg)^{1/2}\bigg) ,
\end{eqnarray}
but additionally require $\sigma(T) / \sigma(T=0)|_{T = 0} = 1$, which amounts 
to $B=1/A-1$. We find $A = 1.39$ to describe the lattice data reasonably 
well.

Using another lattice result, 
$T_c / \sqrt{\sigma(T=0)}\approx 0.71$~\cite{Lucini:2005vg}, allows 
to rewrite Eq. (\ref{EQN666}) according to
\begin{eqnarray}
\label{EQN667} 
\frac{\sigma(T)}{T^2} \ \ = 
\ \ \underbrace{\frac{\sigma(T=0)}{T_c^2}}_{\approx 1.99} 
\bigg(\frac{T_c}{T}\bigg)^2 A \bigg(1-\frac{T}{T_c}\bigg)^{0.63} \bigg(1 + 
            B \bigg(1-\frac{T}{T_c}\bigg)^{1/2}\bigg) .
\end{eqnarray}
Together with (\ref{eqn packing}) this formula relates the density 
of dyons respectively their packing fraction to the temperature ratio
$T / T_c$. Finally, physical units can be introduced using 
$\sigma(T=0) = (440 \, \textrm{MeV})^2$ 
(as we already did in~\cite{Bruckmann:2009nw}) corresponding to 
$T_c = 312 \, \textrm{MeV}$.

Assuming that our dyon gas model provides the correct phenomenological 
value of the string tension we can tell how the density and the
packing fraction have to behave as functions of the temperature 
below $T_c$. For both the limits $T \to T_c$ and $T \to 0$ the density $\rho$ 
tends to zero. Its maximal value $\rho_{max} \approx 0.25 \, T_c \, \sigma(T=0)$ 
is reached at $T \approx 0.65 \,T_c$. 
In physical units we have $\rho_{max} \approx 2~\textrm{fm}^{-3}$. The
packing fraction $f_P$ for our model diverges for 
$T \to 0$ and tends to zero for $T \to T_c$. The latter behavior can be 
interpreted such that the diluteness assumption 
applies best near the phase transition, but becomes more and more 
violated for low temperatures. This problem, however, is well-known to 
occur also for the instanton liquid model~(see e.g.\ \cite{Schafer:1996wv}).

\subsection{Polyakov loop correlator at arbitrary separation and finite-volume 
effects}
\label{subsec:finite_volume}

In this subsection we numerically evaluate the Polyakov loop correlator from 
Eq.\ (\ref{eqn:correlator_third}) and correspondingly the integrals 
$\f_\pm$ from Eq.\ (\ref{eqn:fpm_rescaled}) at arbitrary quark-antiquark 
separation $d$ and arbitrary volume $V$ (both finite and infinite). 
This allows to investigate finite-volume effects. As a by-product we will 
confirm the linear behavior for infinite volume and large separations, 
Eqs.\ (\ref{eqn:free_infvolume},\ref{eqn:string_infvolume}).

To perform the numerical integration efficiently, we split the integrals in 
two regions, $S \equiv S^3_{\tilde{R} T}$, a ball of radius $\tilde{R} < R$, 
and its complement $\bar{S} \equiv S^3_{R T} - S$:
\begin{eqnarray}
f_\pm \ \ = \ \ \int_S d\mathbf{s} \, 
    \exp\bigg(\frac{i}{2} \bigg(\frac{1}{|\mathbf{r} T - \mathbf{s}|} \pm 
    \frac{1}{|\mathbf{r}' T - \mathbf{s}|}\bigg)\bigg) + 
    \int_{\bar{S}} d\mathbf{s} \, \exp\bigg(\frac{i}{2} 
    \bigg(\frac{1}{|\mathbf{r} T - \mathbf{s}|} \pm \frac{1}{|\mathbf{r}' T - 
    \mathbf{s}|}\bigg)\bigg) .
\end{eqnarray}

The integral over $S$ can be solved numerically with standard methods, 
e.g.\ ordinary Monte Carlo sampling, because both the region of integration 
and the integrand are finite. By introducing spherical coordinates it can 
even be reduced to a 2-dimensional integral:
\begin{eqnarray}
\nonumber & & \hspace{-0.7cm} 
f_{\pm,S} \ \ = 
\ \ \int_S d\mathbf{s} \, \exp\bigg(\frac{i}{2} \bigg(\frac{1}{|\mathbf{r} T -
 \mathbf{s}|} \pm \frac{1}{|\mathbf{r}' T - \mathbf{s}|}\bigg)\bigg) \\
\nonumber & & \ \ \ 
= \ \ 2 \pi \int_0^{\tilde{R} T} ds \, s^2 \int_0^\pi d\theta \, \sin\theta 
~ \exp\bigg(\frac{i}{2} \bigg(
\frac{1}{D_+(s,\theta,dT)} \pm 
\frac{1}{D_-(s,\theta,dT)}
\bigg)\bigg) ,
\end{eqnarray}
with $D_{\pm}$ according to Eq. (\ref{eqn:abs_values}).
For $R \rightarrow \infty$ the integrals over $\bar{S}$ exhibit infinities, 
which need to be subtracted, before a numerical treatment is possible. 
For finite but large $R$ this subtraction is essential for an efficient 
computation of the integrals. To exhibit the infinities, we expand in powers 
of $1/s$:
\begin{eqnarray}
\nonumber 
& & \hspace{-0.7cm} 
f_{+,\bar{S}} \ \ = 
\ \ \int_{\bar{S}} d\mathbf{s} \, \exp\bigg(\frac{i}{2} \bigg(\frac{1}{|\mathbf{r} T -
 \mathbf{s}|} + \frac{1}{|\mathbf{r}' T - \mathbf{s}|}\bigg)\bigg) \\
\nonumber & &  \ \ \ = 
\ \ 2 \pi \int_{\tilde{R} T}^{R T} ds \, s^2 \int_0^\pi d\theta \, \sin\theta \\
\nonumber & & \hspace{1cm} 
\bigg(1+{\frac {i}{s}}  -  {\frac {1}{2{s}^{2}}}  +  
{\frac {i \left( -3\,(dT)^2+9\,(dT)^{2}\cos^{2}\theta-4 \right) }{24{s}^{3}}}+
\mathcal{O}(1/s^4)\bigg)\bigg) \\
& &  \ \ \  = \ \ V(\bar{S}) + \Lambda + \textrm{finite} \\
\nonumber & & \hspace{-0.7cm} 
f_{-,\bar{S}} \ \ = 
\ \ \int_{\bar{S}} d\mathbf{s} \, \cos\bigg(\frac{1}{2|\mathbf{r} T - \mathbf{s}|} - 
\frac{1}{2|\mathbf{r}' T - \mathbf{s}|}\bigg) \ \ = 
\ \ \int_{\bar{S}} d\mathbf{s} \, \cos\bigg(\frac{1}{2 s} \bigg(\frac{d T s_z}{s^2} 
+ \mathcal{O}(1/s^3)\bigg)\bigg) \\
 & &  \ \ \ = \ \ 2 \pi \int_{\tilde{R} T}^{R T} ds \, s^2 
\int_0^\pi d\theta \, \sin\theta \bigg(1 + \mathcal{O}(1/s^4)\bigg) \ \ = 
\ \ V(\bar{S}) + \textrm{finite}\,,
\end{eqnarray}
where
\begin{eqnarray}
\Lambda \ &=& 2\pi \,\int _{{\tilde RT}}^{RT}\!{\d s}\,{s}^{2}
          \int _{0}^{\pi }\! \,{\d\theta}\,
           \sin\theta\left( {\frac {i}{s}} - {\frac {1}{2\,{s}^{2}}}+
           {\frac {i \left( -3\,(dT)^{2}+9\,(dT)^{2}\cos^2 \theta-
            4 \right) }{24\,{s}^{3}}} 
           \right)\\
          &=& \frac{2\,\pi}{3} \, 
              \left( - 3\,i{T}^{2} \left( {\tilde R}^{2}-{R}^{2} \right) 
                     + 3\,T \left( {\tilde R}-R \right)
                    +i \ln  \left( {\frac {{\tilde R}}{R}}
              \right)  \right)
\end{eqnarray}
see also (\ref{eqn:Rdep_plus}).
Note that the imaginary part of $f_-$ vanishes, as argued in 
Section~\ref{subsec:confinf_volume}.

The finite parts of the above integrals can be evaluated numerically:
\begin{eqnarray}
\nonumber & & \hspace{-0.7cm} 
f_{+,\bar{S},\textrm{finite}} \ \ = 
\ \ 2 \pi \int_{\tilde RT}^{RT} \d s \, s^2 \int_0^\pi \d\theta \, \sin\theta 
\nonumber
~ \Bigg[\exp\bigg(\frac{i}{2} \bigg(
\frac{1}{D_+(s,\theta,dT)} + 
\frac{1}{D_-(s,\theta,dT)}
\bigg)\bigg) \\
 & &\hspace{2.1cm}
-1-{\frac {i}{s}}
+{\frac {1}{2{s}^{2}}}
-{\frac {i \left( -3\,(dT)^2+9\,(dT)^2\cos^{2}\theta-4 \right) }{24{s}^{3}}}\Bigg]\\
\nonumber & & \hspace{-0.7cm} 
f_{-,\bar{S},\textrm{finite}} \ \ =  
\ \ 2 \pi \int_{\tilde{R} T}^{R T} ds \, s^2 \int_0^\pi d\theta \, \sin\theta 
~ \Bigg[\cos\bigg(
\frac{1}{2 D_+(s,\theta,dT)} - 
\frac{1}{2 D_-(s,\theta,dT)}
\bigg)  - 1 \Bigg] .
\end{eqnarray}

The range of integration of $\int ds$, which extends to infinity in the 
limit $R \rightarrow \infty$, still poses a problem, but can be overcome by 
a change of variables according to
\begin{eqnarray}
ds \, \frac{1}{s^2} \ \ = \ \ dx
\end{eqnarray}
(we have chosen that particular form, because the integrands of 
$f_{+,\bar{S},\textrm{finite}}$ and $f_{-,\bar{S},\textrm{finite}}$ 
are proportional to $1/s^2$ for large $s$.) Consequently,
\begin{eqnarray}
\int_{\tilde{R} T}^s ds' \, \frac{1}{s'^2} \ \ = 
\ \ \int_{x_0}^x dx' \quad \rightarrow \quad s \ \ = 
\ \ \frac{1}{1/\tilde{R} T - x + x_0} .
\end{eqnarray}
For simplicity and without loss of generality we choose $x_0 = 0$ 
in the following. Then
\begin{eqnarray}
\int_{\tilde{R} T}^{R T} ds \, F(s) \ \ = 
\ \ \int_0^{1/\tilde{R} T - 1/R T} dx \, 
 \frac{F(1 / (1/\tilde{R} T - x))}{(1/\tilde{R} T - x)^2} ,
\end{eqnarray}
where the integrand is roughly equally distributed over the finite range 
of integration $0 \leq x \leq 1/\tilde{R} T - 1/R T$, if 
$F(s) \approx \# / s^2$. The final expressions for numerical evaluation are
\begin{eqnarray}
\nonumber & & \hspace{-0.7cm} 
f_{+,\bar{S},\textrm{finite}} \ \ = 
\ \ 2 \pi \int_0^{1/\tilde RT-1/RT} \d x \, s^4 
\int_0^\pi \d\theta \, \sin\theta \nonumber
~ \Bigg[\exp\bigg(\frac{i}{2} \bigg(
\frac{1}{D_+(s,\theta,dT)} + 
\frac{1}{D_-(s,\theta,dT)}
\bigg)\bigg) \\
 & &\hspace{2.1cm}
-1-{\frac {i}{s}}
+{\frac {1}{2{s}^{2}}}
-{\frac {i \left( -3\,(dT)^2+9\,(dT)^2\cos^{2}\theta-4 \right) }{24{s}^{3}}}\Bigg]\\
\nonumber & & \hspace{-0.7cm} 
f_{-,\bar{S},\textrm{finite}} \ \ =  
\ \ 2 \pi \int_0^{1/\tilde{R} T - 1/R T} dx \, s^4 \int_0^\pi d\theta \, \sin\theta 
~ \Bigg[\cos\bigg(
\frac{1}{2 D_+(s,\theta,dT)} - 
\frac{1}{2 D_-(s,\theta,dT)}
\bigg)  - 1 \Bigg] ,
\end{eqnarray}
where $s = 1 / (1/\tilde{R} T - x)$.

In total
\begin{eqnarray}
 & & \hspace{-0.7cm} 
f_+ \ \ = \ \ f_{+,S} + f_{+,\bar{S},\textrm{finite}} + V - V(S) + \Lambda  \\
 & & \hspace{-0.7cm} 
f_- \ \ = \ \ f_{-,S} + f_{-,\bar{S},\textrm{finite}} + V - V(S) .
\end{eqnarray}
The Polyakov loop correlator for maximally non-trivial holonomy is
\begin{eqnarray}
\Big\langle P(\mathbf{r}) P(\mathbf{r}') \Big\rangle \ \ 
= \ \ \frac{1}{2} \exp\bigg(2 K \ln \frac{|f_-|}{V T^3}\bigg) \ - 
\  \frac{1}{2} \exp\bigg(2 K \ln \frac{|f_+|}{V T^3}\bigg)
\end{eqnarray}
(cf.\ Eq.\ (\ref{eqn:correlator_second})). In the limit 
$V \rightarrow \infty$ this equation simplifies to
\begin{eqnarray}
\Big\langle P(\mathbf{r}) P(\mathbf{r}') \Big\rangle \ \ = 
\ \ \frac{1}{2} \exp\bigg(\frac{\rho (f_{-,S} + f_{-,\bar{S},\textrm{finite}} - 
V(S))}{T^3}\bigg)\,.
\label{eqn:correlator_fourth}
\end{eqnarray}
We have performed the remaining integrations numerically 
and show the results below in Figs. \ref{fig:free_energy_separation} 
and \ref{fig:free_energy_limit}.

\section{Ewald's summation method}
\label{sec:ewald}

\subsection{Outline of the method}
\label{subsec:outline}

In the following we briefly summarize Ewald's method. For a more detailed 
presentation we refer to~\cite{Lee:2009}. Our main motivation to use this 
method is to systematically control finite-volume effects in observables,
in particular those contributing to the Polyakov loop in Eq.~(\ref{eqn:Phi}).

The first step in Ewald's method is to mimic the infinite space by sampling 
the physical system restricted to a basic cell, the so-called ``super cell'', 
of spatial volume $L^3$ which will -- for finite density -- contain only a 
finite number of randomly placed dyons. In a second step the space is filled 
with replicas of the super cell shifted by $\mathbf{n} L$, $\mathbf{n} \in 
\mathbb{Z}^3$. Sums over infinitely many dyons in infinite space are replaced 
by sums over these replicas. 

The infinite sum $\Phi$ in the Polyakov loop, Eq.~(\ref{eqn:Phi}), 
is modified to~\footnote{Note that Ewald's method 
is quite general in a sense that it is capable of performing infinite sums of 
arbitrary inverse powers~\cite{Essmann:1995}.}
\begin{eqnarray}
\Phi(\mathbf{r}) \ \ = 
\ \ \sum_{\mathbf{n} \in \mathbb{Z}^3} 
\sum_j \frac{q_j}{|\mathbf{r} - \mathbf{r}_j - \mathbf{n} L|} ,
\label{eqn:sum_over_replica}
\end{eqnarray}
where $j=(i,m)$ is now a superindex running over all dyons and antidyons coming in
equal number ($j$ takes $n_D=2K$ different values).

Naively one might think that such a sum can be approximated by summing over a 
large but finite number of copies of the super cell.
One can show, however, that even though this sum converges, when increasing the 
total volume further and further, it converges to a result that 
differs from the desired infinite sum $\Phi(\mathbf{r})$ 
(cf.\ appendix~\ref{subsec:ewald_vs_other}). The distortion depends on details 
of the charge distribution like surface charges.
Only in the limit $L \rightarrow \infty$ it is expected to be identical to the 
Ewald result.

The third step and key idea of Ewald's method is to split the terms 
$1 / |\mathbf{r} - \mathbf{r}_j - \mathbf{n} L|$ in $\Phi$ in a very specific 
way into an exponentially decaying ``short-range part'' and a smooth 
``long-range part''. While the sum over the terms appearing in the short-range 
part converges exponentially, the sum in the long-range part is carried out in 
Fourier space, where it also converges exponentially. This allows a rather 
efficient computation of the sum in 
Eq. (\ref{eqn:sum_over_replica}) up to arbitrary precision.

In detail the splitting into the short and long-range sum is done in the 
following way:
\begin{eqnarray}
 & & \hspace{-0.7cm} \Phi(\mathbf{r}) \ \ = 
\ \ \Phi^\textrm{short}(\mathbf{r}) + \Phi^\textrm{long}(\mathbf{r}) \\
\label{eqn:EQN378} 
& & \hspace{-0.7cm} \Phi^\textrm{short}(\mathbf{r}) 
\ \ \equiv \ \
\sum_{\mathbf{n} 
\in \mathbb{Z}^3} \sum_j \bigg(1 - 
\textrm{erf}\bigg(\frac{|\mathbf{r} - \mathbf{r}_j - \mathbf{n} L|}%
{\sqrt{2} \lambda}\bigg)\bigg) 
\frac{q_j}{|\mathbf{r} - \mathbf{r}_j - \mathbf{n} L|} \\
\label{eqn:EQN379} 
& & \hspace{-0.7cm} \Phi^\textrm{long}(\mathbf{r}) 
\ \ \equiv \ \
\sum_{\mathbf{n} \in \mathbb{Z}^3} \sum_j 
 \textrm{erf}\bigg(\frac{|\mathbf{r} - \mathbf{r}_j - \mathbf{n} L|}%
{\sqrt{2} \lambda}\bigg) \frac{q_j}{|\mathbf{r} - \mathbf{r}_j - \mathbf{n} L|} ,
\end{eqnarray}
where $~\textrm{erf}~$ denotes the error function.
The physical intuition behind this decomposition becomes clear by computing the 
charge corresponding to this potential, i.e.\ by applying the Laplace operator 
to $\Phi$. Of course, the original potential 
$1 / |\mathbf{r} - \mathbf{r}_j - \mathbf{n} L|$ yields pointlike sources at 
the dyon positions $\mathbf{r}_j + \mathbf{n} L$. The auxiliary term 
$-\textrm{erf}(|\mathbf{r} - \mathbf{r}_j - \mathbf{n} L|/\sqrt{2} \lambda)
/|\mathbf{r} - \mathbf{r}_j - \mathbf{n} L|$ 
yields a continuous charge distribution around the same locations, but with 
Gaussian profile of width $\lambda$ and opposite sign. It is clear that the 
effect of these two charge distributions increasingly cancels in 
$\Phi^\textrm{short}$ with growing distance, actually in an exponential 
manner. 

In $\Phi^\textrm{long}$ the smeared charge generates a non-singular potential 
at the dyon positions. This leads to a convergence in its Fourier transform, 
which is exponential, too:
\begin{eqnarray}
\label{eqn:EQN380} 
\Phi^\textrm{long}(\mathbf{r}) \ \ = 
\ \ \frac{4 \pi}{L^3} \sum_{\mathbf{n} \in \mathbb{Z}^3 \setminus \vec{0}} 
e^{+i \mathbf{k}(\mathbf{n}) \mathbf{r}} ~
\frac{e^{-\lambda^2 \mathbf{k}(\mathbf{n})^2 / 2}}{\mathbf{k}(\mathbf{n})^2} 
\bigg(\sum_{j=1}^N q_j e^{-i \mathbf{k}(\mathbf{n}) \mathbf{r}_j}\bigg) 
\quad , \quad 
\mathbf{k}(\mathbf{n}) \ \ = \ \ \frac{2\pi}{L}\,\mathbf{n} \; .
\end{eqnarray}
The expressions in parentheses are called structure functions, since they 
contain the information about the dyon positions. Note that this expression 
for $\Phi^\textrm{long}$ is correct only if the system is neutral, i.e.\ if 
$\sum_j q_j = 0$. This is the case for the non-interacting dyon model. 
For non-neutral systems $\Phi^\textrm{long}$ diverges.
The free parameter $\lambda$ determines the tradeoff between the long-range 
sum and the short-range sum. While the short-range sum can be evaluated 
rather quickly for small $\lambda$, the opposite is the case for the 
long-range sum. The optimal choice for $\lambda$ is discussed in the 
following section.

\subsection{Performance of Ewald's method}
\label{subsec:performance}

To determine the free energy of a static quark-antiquark pair within the 
non-interacting dyon model, we need to evaluate Polyakov loop correlators.
Doing this in an efficient way amounts to computing $\Phi$ at a set of 
sample points $\mathbf{r}$ distributed on a cubic lattice throughout the 
spatial volume. Let $M$ be the number of sample points. The computational 
costs to evaluate the short-range sum (\ref{eqn:EQN378}) up to any desired 
accuracy are $\mathcal{O}(M \lambda^3)$ assuming that dyons within a 
spherical region around a given sample point can be always identified 
within the same cpu-time (see below, how this can be realized).

Similarly one can read off the computational costs for evaluating the 
long-range sum (\ref{eqn:EQN380}) up to exponential precision. 
The structure functions 
$\sum_j q_j e^{-i \mathbf{k}(\mathbf{n}) \mathbf{r}_j}$ are independent 
of the sample point $\mathbf{r}$ and, therefore, need to be computed 
only once for a given dyon configuration. The number of the required
structure functions is proportional to $V / \lambda^3$, hence the corresponding 
computational costs are of order $\mathcal{O}(V^2 / \lambda^3)$. The time 
needed for the subsequent computation of $\Phi^\textrm{long}$ at all $M$ 
sample points is proportional to $M V / \lambda^3$. 
Consequently, the total computational costs to perform the long-range sum 
are $\mathcal{O}(V^2 / \lambda^3) + \mathcal{O}(M V / \lambda^3)$.

\begin{figure}[htb]
 \vspace{-0.5cm}
 \begin{minipage}{0.49\textwidth}
  \includegraphics[width=0.65\textwidth, angle=-90]{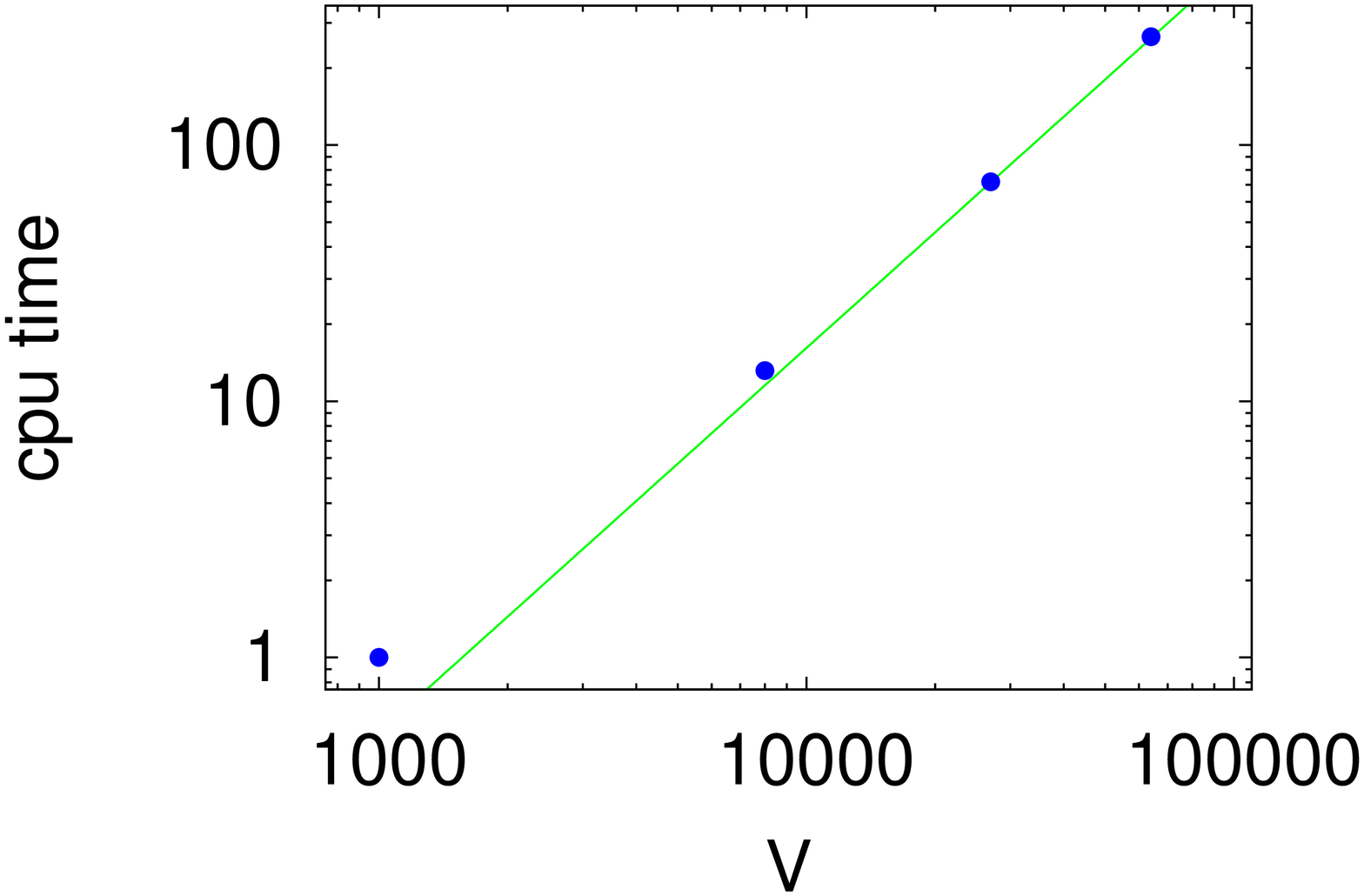}
 \end{minipage}
 \begin{minipage}{0.49\textwidth}
  \includegraphics[width=0.65\textwidth, angle=-90]{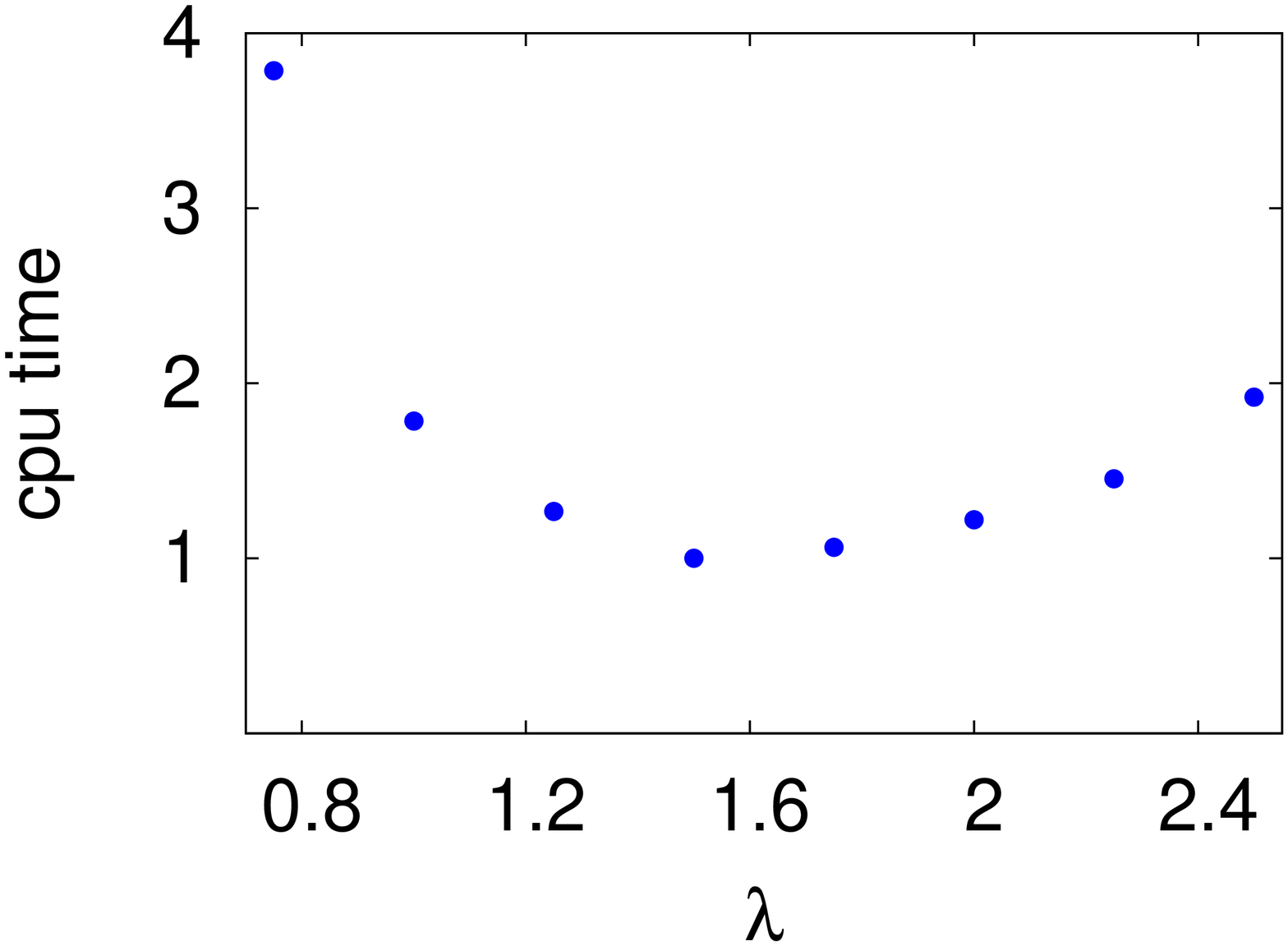}
 \end{minipage}
 \vspace{-1mm}
 \caption{Performance of Ewald's method. \textbf{Left:}
  a log-log plot of
  the computing time needed to evaluate the potential $\Phi$ at $M \propto V$
  sample points per dyon configuration as a function of the spatial volume $V$.
  The density of dyons is $\rho = 1.0$. The vertical axis is labeled such
  that one unit of cpu-time was needed to perform the computation for
  $V = (10.0)^3$ corresponding to a number of dyons $n_D=1000$. $\lambda$
  was chosen according to  $\lambda^3 \propto \sqrt{V}$.
  The straight line with slope $3/2$ illustrates that for large spatial
  volumes/dyon numbers Ewald's method indeed exhibits the expected
  $\mathcal{O}(V^{3/2})$ scaling. \textbf{Right:} the computing time as a
  function of the parameter $\lambda$ (in units of the inverse temperature)
  for $n_D = 8000$ and $V = (20.0)^3$.
  The vertical axis is labeled such, that one unit of cpu-time was needed to
  perform the computation at the optimal value $\lambda_{\rm opt}\approx 1.5$.}
\label{fig:ewald_scaling}
\end{figure}

The computational costs of the short-range sum and of the long-range sum 
depend on $\lambda$ in just the opposite way (as expected). 
One should choose 
$\lambda$ in an optimal way, such that the total computational costs are 
minimized. Obviously the optimal choice for $\lambda$ also depends on $M$. 
Since typically $M \propto V$, as it is the case for our computations, the 
optimal value for $\lambda$ should be chosen according to 
$\lambda^3 \propto \sqrt{V}$. Then the performance of Ewald's method is 
$\mathcal{O}(V^{3/2})$. This behavior has been confirmed numerically, 
cf.\ Fig.~\ref{fig:ewald_scaling}, left panel.

Of course, $\lambda^3 \propto \sqrt{V}$ is only a statement on how to 
increase $\lambda$, when enlarging the spatial volume $V$. 
How to choose $\lambda$ for a given $V$ such that the corresponding 
computing time is minimized, has to be determined by numerical experiment. 
In the right panel of Fig. \ref{fig:ewald_scaling} we show in an exemplary 
plot corresponding to $n_D = 8000$ and $V = (20.0)^3$ the computing time 
needed to calculate the dyon potential $\Phi$ as a function of $\lambda$. 
Obviously, there is an optimal choice for $\lambda$.

Note that in the literature there also exists another version
of the just explained ``classical Ewald method'', the so-called 
``particle mesh Ewald method'' (cf.\ e.g.\ \cite{Essmann:1995}). This 
version is more efficient, when the interaction energy of a system of positive 
and negative charges needs to be computed. However, for our problem at 
hand, the computation of the temporal gauge field $\Phi$,  there is no 
advantage with respect to performance. 
Since it is significantly simpler to implement, we resort to 
the classical Ewald method.

For an efficient computation of the short-range sum 
$\Phi^\textrm{short}(\mathbf{r})$ it is mandatory to determine, which dyons 
are located in a spherical region of given radius $R$ centered around 
$\mathbf{r}$ in $\mathcal{O}(1)$ computer time. To this end we divide the 
supercell into a grid of cubic subcells and generate for each subcell a 
list of the contained dyons. In addition we have implemented a function 
that determines all subcells, which are inside or which intersect the 
surface of the above mentioned ball. Then we call all those subcells for 
the dyons they contain. In this way we do not need to inspect
all the dyons in the supercell and check whether their distance to 
$\mathbf{r}$ is smaller than $R$. Of course, the grid of subcells has to be 
sufficiently fine-grained, to be able to mimic a ball
of radius $R$ with rather small cubes.

\section{Numerical results}
\label{sec:results}

\subsection{Extracting the infinite volume string tension using Ewald's method}
\label{subsec:F_QQ}

We compute the free energy of a static quark-antiquark pair as a function of 
their separation from Polyakov loop correlators as described in 
Section~\ref{sec:dyons}. We keep the dyon density $\rho$ and the temperature 
$T$ fixed and perform computations for various dyon numbers $n_D$, corresponding 
to various spatial volumes $V = n_D / \rho$ of the super cell. The superposition 
of dyon potentials $\Phi$ is calculated by means of the Ewald method as explained 
in Section~\ref{sec:ewald}. We restrict ourselves to maximally non-trivial 
holonomy.

Of course, the resulting free energies are different for different dyon 
numbers, i.e. spatial volumes of the super cell, because of finite-volume effects. 
In particular the dyon potential $\Phi$ is $L$-periodic along the three spatial 
directions, which obviously implies periodic Polyakov loops and loop correlators. 
Therefore, $L$ has to be chosen sufficiently large to ensure that the free energy 
can be determined for quark-antiquark separations of phenomenological interest, 
typically a few $\text{fm}$, without being significantly
distorted due to periodicity.

\begin{figure}[!t]
\centering
\includegraphics[width=0.75\textwidth]{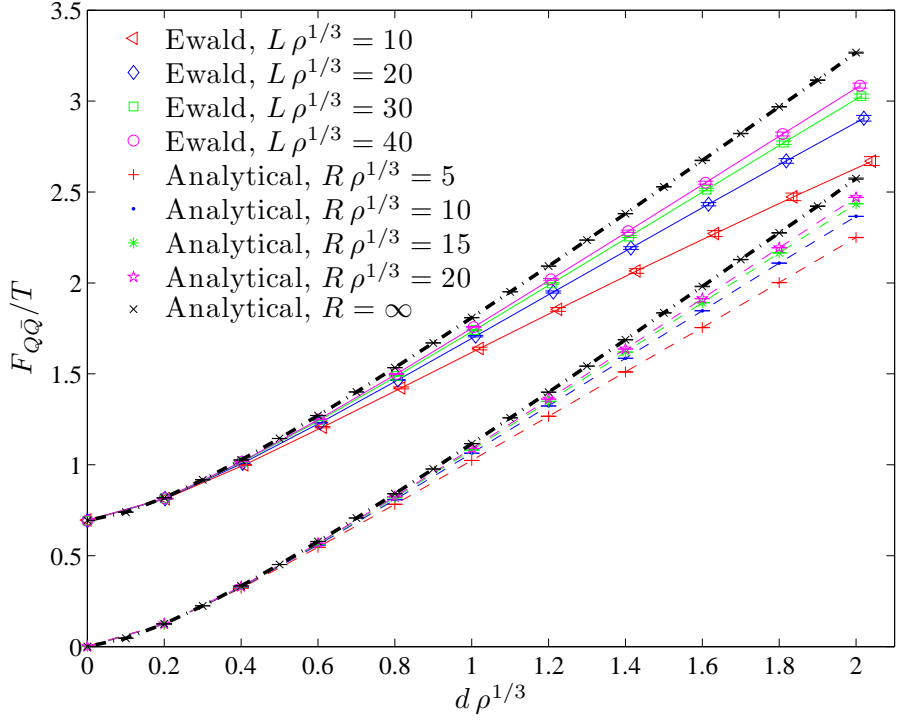}
\caption{Free energy of a static quark-antiquark pair as a function of its 
separation for $\rho / T^3 = 1.0$ and various supercell extensions 
$L \rho^{1/3}$ corresponding do different dyon numbers $n_D$. In addition 
we show the results obtained from a numerical evaluation of the analytic result
at finite and infinite volume. For better visibility the analytic results 
are shifted by $\log{2}$ and therefore the corresponding curves start close 
to the origin.}
\label{fig:free_energy_separation}
\end{figure}
\begin{table}[!b]
\begin{center}
\begin{tabular}{|r|r|r|}
 \hline
\vspace{-0.4cm} & & \\
 $n_D$ & $L \rho^{1/3}$ & \# configurations \\
\vspace{-0.4cm} & & \\
 \hline
\vspace{-0.4cm} & & \\
   $1000$ & $10$ & $1600$ \\ 
   $8000$ & $20$ &  $800$ \\
  $27000$ & $30$ &  $120$ \\ 
  $64000$ & $40$ &   $90$ \\ 
 $125000$ & $50$ &   $60$\vspace{-0.4cm} \\
 & & \\
 \hline
\end{tabular}
\end{center}
\caption{Number of random dyon configurations used for every 
simulation at fixed dyon number $n_D$ or dimensionless length 
of the volume in which the dyon positions are sampled, 
$L\rho^{1/3}$,
respectively.}
\label{tab:numConfs}
\end{table}

\begin{figure}[htb]
\includegraphics[width=\textwidth]{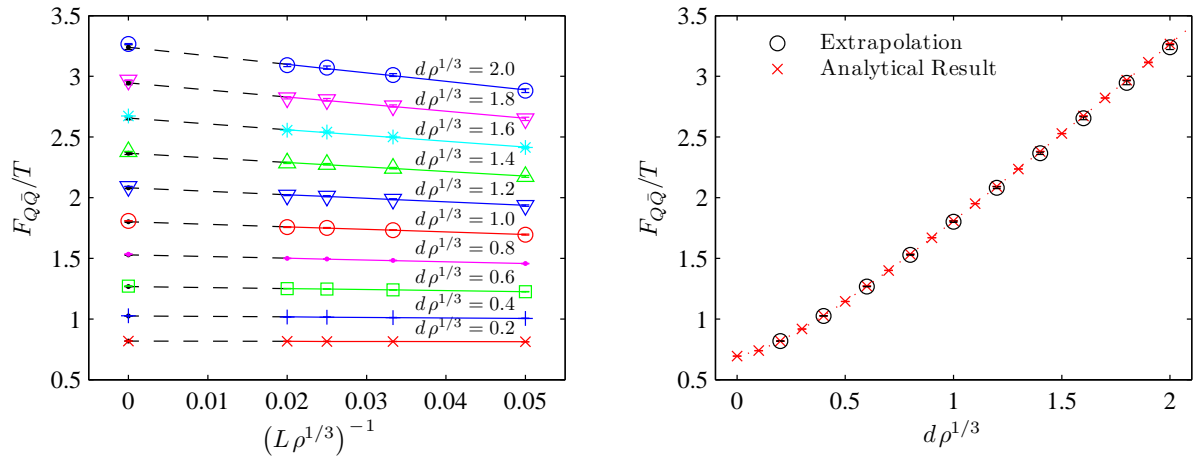}
\caption{Demonstration of the infinite volume limit. 
\textbf{Left:} Static free energy 
for ten distances as a function of the inverse supercell extensions 
$L \rho^{1/3}$ (corresponding to different dyon numbers $n_D$)
and its extrapolation to infinite volume. 
\textbf{Right:} The latter compared to the analytic result, 
Eq.~(\protect\ref{eqn:correlator_fourth}), for arbitrary distances in 
infinite volume.
}
\label{fig:free_energy_limit}
\end{figure}

In Fig.~\ref{fig:free_energy_separation} we show quark-antiquark free energies 
for $\rho / T^3 = 1.0$ and dyon numbers 
$1000 \leq n_D \leq 125000$ 
(corresponding to $10.0 \leq L \rho^{1/3} \leq 50.0$) as functions of the 
quark-antiquark separation $d \rho^{1/3}$. We also show the analytic results for 
finite and infinite volume in this plot. Note that we express lengths in units 
of $\rho^{1/3}$, which is the average dyon separation in a random dyon gas. 
The number of dyon configurations used for each dyon number $n_D$ is listed 
in Table~\ref{tab:numConfs}. It can be seen that the free energies converge, 
when increasing $n_D$ (or equivalently $L \rho^{1/3}$). This allows an 
extrapolation to infinite volume. In the left panel of Fig.~\ref{fig:free_energy_limit} 
we show linear extrapolations of the finite-volume static free energy 
to infinite volume for a number of quark-antiquark separations. We also compare 
the results of the extrapolation to the analytically obtained free energy at 
infinite volume in the right panel of Fig.~\ref{fig:free_energy_limit}. As can 
be seen, analytic and extrapolated results nicely agree within errors.

Let us point out that there are other methods of obtaining an 
infinite volume result numerically without employing Ewald's 
summation method.

An obvious method is a straightforward superposition of dyon 
potentials in a finite cubic box of size $L^3$, that we call 
\textit{dyon sampling volume}. 
We have used this method in a previous publication 
\cite{Bruckmann:2009nw}, to which we refer for further details. 
Note that there is no exact translational invariance anymore, 
in contrast to when using periodic boundary conditions via 
Ewald's method. To keep finite-volume effects at a tolerable level, 
we have to restrict the evaluation of Polyakov loops to a spatial 
subvolume sufficiently far away from the boundary of the dyon 
volume. We will call this subvolume \textit{field sampling volume}. 
It is centered inside the dyon sampling volume and has extension 
$l \leq L$. On the one hand, finite-volume effects are expected 
to be negligible for sufficiently small $l$. On the other hand, 
however, decreasing $l$ reduces the available information per 
dyon configuration and, therefore, reduces statistical accuracy. 
In practice one would need to identify plateaus in the observables 
as functions of $l$.
An obvious and major drawback of proceeding in such a way is that 
one needs to extrapolate in two parameters, the extension $l$ of the 
sampling volume and the extension $L$ of the dyon volume. Clearly this 
is technically more complicated, than what has to be done using 
Ewald's method, where the only parameter subject to extrapolation is 
the extension of the supercell $L$.

One can also think of evaluating just one Polyakov loop correlator in 
the center of the volume of each random configuration. We should point 
out that this is not really feasible if there are interactions, since 
a significantly larger statistics is needed when no volume averaging is 
done. For the non-interacting case it is applicable and therefore 
worth being mentioned. 

\section{Summary and outlook}
\label{sec:summary}

In this work we have shown analytically that a non-interacting random 
dyon gas leads to a linearly rising free energy of a static quark-antiquark 
pair as a function of the distance in between. Correspondingly the string 
tension $\sigma$ turned out proportional to the ratio of the density and the 
temperature, i.e. to $\rho/T$, cf.\ Eq.~(\protect\ref{eqn:string_infvolume}).
We were able to present explicit formulae for 
arbitrary distance and for finite volume with certain integrals left to be 
evaluated numerically. We convinced ourselves that the dependence on the
holonomy drops out in the infinite volume limit. This reflects the fact 
that -- concerning the Polyakov loop and its correlator -- 
the model is able to describe only the confinement phase. For the deconfinement 
transition as well as for the deconfinement phase, where 
the center symmetry becomes broken, the model should be altered 
taking into account that dyons with opposite charge should be statistically 
weighted differently.

We emphasize, that our analytical approach is specific for the non-interacting 
case. For the interacting case it will not be applicable without approximations, 
and in the first instance we will have to rely on numerical simulations.
Strong finite-size effects of the naive treatment with finite boxes containing 
the dyon sources have led us to employ a numerical method well-known in the 
physics of a three-dimensional Coulomb plasma, the Ewald summation method. 
We convinced ourselves that this method will be applicable also to the 
more realistic interacting dyon gas.      
 
Indeed, we have demonstrated, how Ewald's summation method can be used to deal 
with long-range objects also in field theory, in our case with random ensembles 
of dyon constituents. In this semiclassically motivated model we have computed 
the local Polyakov loop, and from its correlator we have extracted the string 
tension, the main observable characterizing confinement/deconfinement at 
finite temperature.

The Polyakov loop is a function of an infinite sum of Coulomb contributions 
of dyons with both signs of charge (cf.\ Eq.~(\ref{eqn:Phi})). According to 
Ewald's method we have decomposed this sum into short-range and long-range 
parts, Eqs.~(\ref{eqn:EQN378}) and (\ref{eqn:EQN380}), and have optimized 
the width $\lambda$ of the auxiliary Gaussian charge cloud governing the 
strength of the exponential convergence of both parts. 

Figs.~\ref{fig:free_energy_separation} and~\ref{fig:free_energy_limit} show
our main results, the free energy of a quark-antiquark pair as a function 
of its separation, for various extensions of the (periodically repeated)
supercell volume, but fixed dyon density. These figures also 
demonstrate that the straightforward extrapolation to infinite 
supercell volume is a valid procedure to obtain results for an infinite 
non-interacting system of dyons. In this limit the Polyakov loop correlator 
behaves as expected: it decays exponentially toward larger quark-antiquark 
separations. The corresponding string tension can be read off unambiguously
(and used to fix the physical scale of this model).
 
We have discussed the advantages of Ewald's periodic summation over 
methods that at finite volumes measure observables only in subvolumes: 
it keeps translational invariance and the infinite volume limit amounts 
to extrapolating just one parameter.

The applicability of the numerical method we have used is not restricted to 
a non-interacting dyon ensemble and/or to $SU(2)$. 
Dyon fields in higher gauge groups decay with the distance in the same Coulombic 
manner, just possessing different color structures. Several other ingredients 
of dyon models contain Coulomb tails, too, like the interaction of dyons via 
the action or their moduli space metric. 
Furthermore, spatial Wilson loops (providing an area law decay 
with magnetic screening persistent also in the deconfined phase) 
can -- with the help of Stokes' theorem and based on (anti)selfduality -- 
be represented as area integrals over the normal component of the gradient 
of the same infinite sum. 

The ability to perform a controlled infinite volume extrapolation (with a 
single remaining parameter $L$, the extension of the periodically continued 
spatial volume) is even more important in more complicated systems. 
An ensemble of random dyons could be treated easily with up to $10^5$ dyons. 
Interacting dyon ensembles are numerically much more expensive such that the 
reduction to a smaller number of dyons most likely cannot be avoided. 
Then finite-volume effects might become a limiting factor.
Consequently, Ewald's summation method seems to become indispensable, 
however, in form of the particle mesh Ewald method, which is more efficient 
than the classical Ewald method, when computing dyon interactions.

Finally, one could think about applying Ewald's method to more complicated 
objects in gauge theory, whose corresponding fields are also of long-range 
nature, such as merons or regular gauge instantons~\cite{Lenz:2003jp,Lenz:2007st} 
and generalizations thereof~\cite{Wagner:2006qn,Szasz:2008qk}.

\appendix
 
\section{Calculation of some integrals}
\label{app:integrals}

We derive the following results for the integrals $g_-^{(2)}$ and 
$g_-^{({\rm res})}$ involved in Polyakov loop correlators in 
Section~\ref{subsec:confinf_volume},
\begin{eqnarray}
\int_{\mathbb{R}^3} d\mathbf{s} ~
 \bigg(\frac{1}{|\mathbf{u}-\mathbf{s}|} - 
\frac{1}{|\mathbf{u'} -\mathbf{s}|}\bigg)^2
  &=&4\pi|\mathbf{u}-\mathbf{u'}|\\
0 < 
\int_{\mathbb{R}^3} d\mathbf{s} \,
\bigg\{\cos\bigg(\frac{1}{|\mathbf{u} -\mathbf{s}|} - 
\frac{1}{|\mathbf{u'} -\mathbf{s}|}\bigg)&-&1
+\frac{1}{2}\bigg(\frac{1}{|\mathbf{u}-\mathbf{s}|} - 
\frac{1}{|\mathbf{u'} -\mathbf{s}|}\bigg)^2\bigg\}
<\mbox{const.}
\end{eqnarray}
In the first integral we use the well-known Fourier representation of 
the Coulomb potential
\begin{eqnarray}
 \frac{1}{|\mathbf{s}|}
=\frac{1}{(2\pi)^3}\int \!d\mathbf{p}\, \frac{4\pi}{p^2}
e^{i\mathbf{p}\mathbf{s}}
\end{eqnarray}
to calculate
\begin{eqnarray}
 &&\int\! d\mathbf{s} 
  \bigg(\frac{1}{|\mathbf{u}-\mathbf{s}|} - \frac{1}{|\mathbf{u'} 
  -\mathbf{s}|}\bigg)^2\nonumber\\
 &=&\frac{1}{4\pi^4}\int\! d\mathbf{s}\, 
  \int\! d\mathbf{p}\,d\mathbf{q}\,
  \frac{1}{p^2 q^2}\bigg(
  e^{i\mathbf{p}(\mathbf{s}-\mathbf{u})}-
  e^{i\mathbf{p}(\mathbf{s}-\mathbf{u'})}
  \bigg)\bigg(
  e^{i\mathbf{q}(\mathbf{s}-\mathbf{u})}-e^{i\mathbf{q}(\mathbf{s}-\mathbf{u'})}
  \bigg)\nonumber\\
 &=&\frac{1}{4\pi^4}\int\! d\mathbf{p}\,d\mathbf{q}\,
  \frac{1}{p^2 q^2}\delta(\mathbf{p}+\mathbf{q})
  \bigg(e^{-i\mathbf{p}\mathbf{u}}-e^{-i\mathbf{p}\mathbf{u'}}
  \bigg)\bigg(
  e^{-i\mathbf{q}\mathbf{u}}-e^{-i\mathbf{q}\mathbf{u'}}
  \bigg)\nonumber\\
 &=&\frac{2}{\pi}\int\! d\mathbf{p}\,\frac{1}{p^4}
  \bigg(2-2\cos\left(\mathbf{p}(\mathbf{u}-\mathbf{u'})\right)\bigg)
  =8\int_0^\infty \!dp\, \frac{1}{p^2}
  \int_0^\pi \!d\theta\sin\theta
  \bigg(1-\cos\left(p|\mathbf{u}-\mathbf{u'}|\cos\theta\right)\bigg)\nonumber\\
 &=&8\int_0^\infty\! dp\, \frac{1}{p^2}
  \bigg(2-2\,\frac{\sin\left(p|\mathbf{u}-\mathbf{u'}|\right)}
  {p|\mathbf{u}-\mathbf{u'}|}\bigg)
  =4\pi|\mathbf{u}-\mathbf{u'}|
\end{eqnarray}
The Laplace operator can be used to check this result. 
Acting with respect to $\mathbf{u'}$ and $\mathbf{u}$ on the left 
hand side we obtain (from the mixed term in the integrand)
\begin{eqnarray}
 \Delta_{u}\Delta_{u'}\int\! d\mathbf{s}\,
 \bigg(\frac{1}{|\mathbf{u}-\mathbf{s}|} - 
\frac{1}{|\mathbf{u'} -\mathbf{s}|}\bigg)^2
 &=& -2 \int\! d\mathbf{s}\, (-4\pi)^2\delta(\mathbf{u}-\mathbf{s})
\delta(\mathbf{u'}-\mathbf{s})\nonumber\\
 &=&-32\pi^2\delta(\mathbf{u}-\mathbf{u'})
\end{eqnarray}
On the right hand side it gives the same since
\begin{eqnarray}
 \Delta_{u}\Delta_{u'}\,4\pi|\mathbf{u}-\mathbf{u'}|=
\Delta_{u}\,\frac{8\pi}{|\mathbf{u}-\mathbf{u'}|}=
-32\pi^2\delta(\mathbf{u}-\mathbf{u'})\,.
\end{eqnarray}

The integrand of the second integral $\cos x -1 + x^2/2 \equiv h(x)$ is 
positive, which proves the first inequality. For the second inequality
we split $\mathbf{s}$-space into two half-spaces, 
$|\mathbf{u}-\mathbf{s}|\lessgtr|\mathbf{u'}-\mathbf{s}|$, separated 
by the midplane between the two points $\mathbf{u}$ and $\mathbf{u'}$. 
The integral over each half-space gives half of the full integral 
and thus we can specify to one of them, e.g.\ where
\begin{eqnarray}
 0 \leq \frac{1}{|\mathbf{u} -\mathbf{s}|} - 
\frac{1}{|\mathbf{u'} -\mathbf{s}|} < \frac{1}{|\mathbf{u} -\mathbf{s}|}
\end{eqnarray}
holds. Since the integrand $h(x)$ is monotonically increasing 
for $x>0$, we obtain an upper bound
\begin{eqnarray}
 &&\int_{\mathbb{R}^3} d\mathbf{s} \,
  \bigg\{\cos\bigg(\frac{1}{|\mathbf{u} -\mathbf{s}|} - 
  \frac{1}{|\mathbf{u'} -\mathbf{s}|}\bigg)-1
  +\frac{1}{2}\bigg(\frac{1}{|\mathbf{u}-\mathbf{s}|} - 
  \frac{1}{|\mathbf{u'} -\mathbf{s}|}\bigg)^2\bigg\}
  \nonumber\\
 &=&2 \int_{|\mathbf{u}-\mathbf{s}|\leq|\mathbf{u'}-\mathbf{s}|}
d\mathbf{s} \,
  \bigg\{\cos\bigg(\frac{1}{|\mathbf{u} -\mathbf{s}|} - 
  \frac{1}{|\mathbf{u'} -\mathbf{s}|}\bigg)-1
  +\frac{1}{2}\bigg(\frac{1}{|\mathbf{u}-\mathbf{s}|} - 
  \frac{1}{|\mathbf{u'} -\mathbf{s}|}\bigg)^2\bigg\}
  \nonumber\\
 &<&2 \int_{|\mathbf{u}-\mathbf{s}|\leq|\mathbf{u'}-\mathbf{s}|}
d\mathbf{s} \,
\bigg\{\cos\frac{1}{|\mathbf{u} -\mathbf{s}|}-1
+\frac{1}{2|\mathbf{u}-\mathbf{s}|^2}\bigg\}
\end{eqnarray}

Due to the positivity of the integrand, we can extend the latter integral 
back to the full space and by virtue of translational invariance
put $\mathbf{u}=0$ obtaining another bound
\begin{eqnarray}
 &&\int_{\mathbb{R}^3} d\mathbf{s} \,
\bigg\{\cos\bigg(\frac{1}{|\mathbf{u} -\mathbf{s}|} - 
\frac{1}{|\mathbf{u'} -\mathbf{s}|}\bigg)-1
+\frac{1}{2}\bigg(\frac{1}{|\mathbf{u}-\mathbf{s}|} - 
\frac{1}{|\mathbf{u'} -\mathbf{s}|}\bigg)^2\bigg\}
\nonumber\\
 &<&2\cdot4\pi\int_0^\infty \!ds\, s^2 \bigg\{\cos\frac{1}{s}-1
+\frac{1}{2s^2}\bigg\}=\frac{2\pi^2}{3}
\end{eqnarray}
independently of $|\mathbf{u}-\mathbf{u'}|$, by which
we have proven the second inequality.

\section{Ewald's sum compared with summing over a finite array of 
supercells}
\label{subsec:ewald_vs_other}

Ewald's method amounts to summing over infinitely many copies of 
the cubic spatial volume $L^3$ called supercell. 
An alternative approach is to truncate this sum after a large 
but finite number of copies in every spatial direction 
$\pm x$, $\pm y$ and $\pm z$. The corresponding dyon potential 
obtained by summing over $(2n+1)^3$ copies of the supercell is then
\begin{eqnarray}
\label{eqn:EQN609} \Phi^\textrm{finite sum}(\mathbf{r}) \ \ = 
\ \ \sum_{n_x = -n}^{+n} \sum_{n_y = -n}^{+n} \sum_{n_z = -n}^{+n} 
\sum_j \frac{q_j}{|\mathbf{r} - \mathbf{r}_j - \mathbf{n} L|} ,
\end{eqnarray}
where $\mathbf{r} , \mathbf{r}_j \in [-L/2,+L/2]^3$
and $\mathbf{n} = (n_x,n_y,n_z)$. 
One might expect that, when $n$ is chosen sufficiently large, 
the Ewald result, denoted by $\Phi^\textrm{Ewald}$, and 
$\Phi^\textrm{finite sum}$ become arbitrarily close. In this section 
we explain that this is not the case, i.e.\ that even though 
$\Phi^\textrm{finite sum}$ converges, when increasing $n$, it will 
in general differ from $\Phi^\textrm{Ewald}$.

The difference between the two approaches is
\begin{eqnarray}
\nonumber & & \hspace{-0.7cm} \Delta \Phi(\mathbf{r}) \ \ = 
\ \ \Phi^\textrm{Ewald}(\mathbf{r}) - \Phi^\textrm{finite sum}(\mathbf{r}) 
\\
\label{eqn:EQN654} 
& & = \ \ \sum_{n_x \in \mathbb{Z} \setminus \{ -n,\ldots,+n \}} 
\sum_{n_y \in \mathbb{Z} \setminus \{ -n,\ldots,+n \}} 
\sum_{n_z \in \mathbb{Z} \setminus \{ -n,\ldots,+n \}} 
\sum_j^{n_D}
\frac{q_j}{|\mathbf{r} - \mathbf{r}_j - \mathbf{n} L|} .
\end{eqnarray}
In the following we demonstrate by means of a simple example 
that $\Delta \Phi(\mathbf{r}) \neq 0$ in general. To this end consider 
$n_D = 2$, a dyon ($q_1 = +1$) at position $\mathbf{r}_1 = (-d/2,0,0)$ 
and an antidyon ($q_2 = -1$) at position $\mathbf{r}_2 = (+d/2,0,0)$.

For $d = L$ dyons and antidyons in (\ref{eqn:EQN654}) cancel exactly 
with exception of antidyons/dyons located on planes at 
$(\mp (n+1/2),n_y,n_z) L$, $n_y,n_z \in \mathbb{Z}$. Since the dyon 
potential is identical to the potential of an electric charge in 
classical electrostatics, the situation is reminiscent to that of a 
uniformly polarized cubic dielectric with volume $((2n+1) L)^3$. 
For $n \gg 1$ the discrete charges can be approximated by the surface 
charge density $\sigma = \pm 4 \pi / L^2$ at the two opposite sides 
$x = \pm (n+1/2) L$. 

For $d < L$ the dyon and antidyon potentials only partly cancel 
resulting in a reduced surface charge density $\sigma = \pm 4 \pi d / L^3$.

\begin{figure}[htb]
\begin{center}
\includegraphics[width=0.95\textwidth,angle=0]{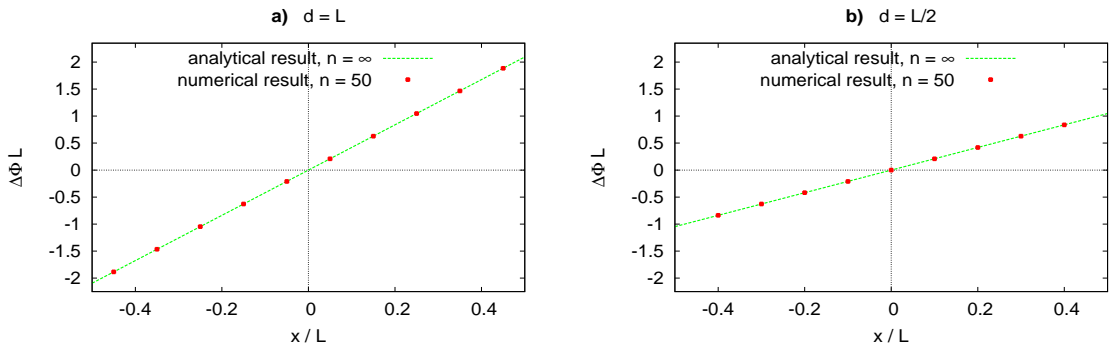}
\caption{$L \Delta \Phi$ as a function of $x/L$ ($y=z=0$) for a dyon at 
$(-d/2,0,0)$ and an antidyon at $(+d/2,0,0)$ 
(cf.\ text for details). \textbf{a)~}$d = L$. \textbf{b)~}$d = L/2$.}
\label{fig:FIG001}
\end{center}
\end{figure}

For $n \gg 1$ the difference $\Delta \Phi$ is given by
\begin{eqnarray}
\nonumber & & \hspace{-0.7cm} 
\nabla (\Delta \Phi(\mathbf{r})) \ \ = 
\ \ -2 \int_{-(n+1/2) L}^{+(n+1/2) L} dy \, \int_{-(n+1/2) L}^{+(n+1/2) L} dz \, 
\frac{d (\mathbf{r} - ((n+1/2)L,y,z)}{4 \pi L^3 |\mathbf{r} - ((n+1/2)L,y,z)|^3} 
\\
 & & = \ \ \frac{4 \pi d}{3 L^3} \Big(\mathbf{e}_x + \mathcal{O}(1/n)\Big) ,
\end{eqnarray}
i.e.\
\begin{eqnarray}
\Delta \Phi(\mathbf{r}) \ \ = \ \ \frac{4 \pi d}{3 L^3} x 
\Big(1 + \mathcal{O}(1/n)\Big) .
\end{eqnarray}
In Fig.~\ref{fig:FIG001} 
we show that this analytical result is accurately 
reproduced by our numerical implementation of Ewald's method and the 
finite sum (\ref{eqn:EQN609}) using $n = 50$.

For a larger number of dyons with arbitrary positions $\Delta \Phi$ is, 
of course, rather hard to estimate analytically. The physical picture, 
however, will remain the same: like in a polarized dielectric surface 
charges will cause a difference between $\Phi^\textrm{Ewald}$ and 
$\Phi^\textrm{finite sum}$. Only in the limit $n_D \rightarrow \infty$ 
corresponding to $L \rightarrow \infty$ both approaches are expected 
to become identical.

In principle both approaches can be used to simulate dyon ensembles, 
since, after appropriately extrapolating the dyon number 
$n_D \rightarrow \infty$ (or alternatively $L \rightarrow \infty$), 
one should obtain the same correct infinite volume result. We consider, 
however, Ewald's method to be superior, because in this approach the 
spatial volume is translationally invariant. This allows to maximally 
exploit a given dyon gauge field configuration by evaluating observables 
throughout the whole spatial volume. In contrast to that, translational 
invariance is broken when truncating the sum over copies of the super cell. 
Observables must only be evaluated in regions, where this breaking 
is sufficiently mild. Each observable requires to determine a corresponding 
region of sufficiently mild finite volume effects. Moreover, one has to 
assure that the associated systematic is removed by the infinite volume 
extrapolation.

\section*{Acknowledgments}
The authors express their gratitude for financial support by the 
German Research Foundation (DFG) with various grants: F.B.\ with grant 
BR 2872/4-2, S.D.\ by the corroborative research center SFB/TR9, E.-M.I.\ 
and M.M.-P.\ with grant Mu 932/6-1, as well as M.W.\ by the Emmy Noether 
Programme with grant WA 3000/1-1.

\bibliographystyle{h-physrev}

\begin{thebibliography}{10}

\bibitem{Callan:1977gz}
C.~G. Callan~Jr, R.~F. Dashen, and D.~J. Gross,
\newblock Phys. Rev. {\bf D17}, 2717 (1978).

\bibitem{Callan:1978bm}
C.~G. Callan~Jr, R.~F. Dashen, and D.~J. Gross,
\newblock Phys. Rev. {\bf D19}, 1826 (1979).

\bibitem{Belavin:1975fg}
A.~A. Belavin, A.~M. Polyakov, A.~S. Shvarts, and Y.~S. Tyupkin,
\newblock Phys. Lett. {\bf B59}, 85 (1975).

\bibitem{Schafer:1996wv}
T.~Sch{\"a}fer and E.~V. Shuryak,
\newblock Rev. Mod. Phys. {\bf 70}, 323 (1998), hep-ph/9610451.

\bibitem{Diakonov:2002fq}
D.~Diakonov,
\newblock Prog. Part. Nucl. Phys. {\bf 51}, 173 (2003), hep-ph/0212026.

\bibitem{Kraan:1998pm}
T.~C. Kraan and P.~van Baal,
\newblock Nucl. Phys. {\bf B533}, 627 (1998), hep-th/9805168.

\bibitem{Kraan:1998sn}
T.~C. Kraan and P.~van Baal,
\newblock Phys. Lett. {\bf B435}, 389 (1998), hep-th/9806034.

\bibitem{Lee:1998bb}
K.-M. Lee and C.-H. Lu,
\newblock Phys. Rev. {\bf D58}, 025011 (1998), hep-th/9802108.

\bibitem{Ewald:1921}
P.~Ewald,
\newblock Ann. Phys. {\bf 369}, 253 (1921).

\bibitem{Harrington:1978ve}
B.~J. Harrington and H.~K. Shepard,
\newblock Phys. Rev. {\bf D17}, 2122 (1978).

\bibitem{Gerhold:2006sk}
P.~Gerhold, E.-M. Ilgenfritz, and M.~M{\"u}ller-Preussker,
\newblock Nucl. Phys. {\bf B760}, 1 (2007), hep-ph/0607315.

\bibitem{Diakonov:2004jn}
D.~Diakonov, N.~Gromov, V.~Petrov, and S.~Slizovskiy,
\newblock Phys. Rev. {\bf D70}, 036003 (2004), hep-th/0404042.

\bibitem{Kraan:1998pn}
T.~C. Kraan,
\newblock Commun. Math. Phys. {\bf 212}, 503 (2000), hep-th/9811179.

\bibitem{Diakonov:2005qa}
D.~Diakonov and N.~Gromov,
\newblock Phys. Rev. {\bf D72}, 025003 (2005), hep-th/0502132.

\bibitem{Bruckmann:2009pa}
F.~Bruckmann, E.-M. Ilgenfritz, B.~Martemyanov, and B.~Zhang,
\newblock Phys. Rev. {\bf D81}, 074501 (2010), 0912.4186.

\bibitem{Manton:1985hs}
N.~S. Manton,
\newblock Phys. Lett. {\bf B154}, 397 (1985).

\bibitem{Gibbons:1986df}
G.~W. Gibbons and N.~S. Manton,
\newblock Nucl. Phys. {\bf B274}, 183 (1986).

\bibitem{Gibbons:1995yw}
G.~W. Gibbons and N.~S. Manton,
\newblock Phys. Lett. {\bf B356}, 32 (1995), hep-th/9506052.

\bibitem{Diakonov:2007nv}
D.~Diakonov and V.~Petrov,
\newblock Phys. Rev. {\bf D76}, 056001 (2007), 0704.3181.

\bibitem{Diakonov:2009jq}
D.~Diakonov,
\newblock Nucl.Phys.Proc.Suppl. {\bf 195}, 5 (2009), 0906.2456.

\bibitem{Polyakov:1976fu}
A.~M. Polyakov,
\newblock Nucl. Phys. {\bf B120}, 429 (1977).

\bibitem{Bruckmann:2009nw}
F.~Bruckmann, S.~Dinter, E.-M. Ilgenfritz, M.~M{\"u}ller-Preussker, and
  M.~Wagner,
\newblock Phys. Rev. {\bf D79}, 116007 (2009), 0903.3075.

\bibitem{Bornyakov:2008im}
V.~Bornyakov, E.-M. Ilgenfritz, B.~Martemyanov, and M.~M{\"u}ller-Preussker,
\newblock Phys.Rev. {\bf D79}, 034506 (2009), 0809.2142.

\bibitem{Bruckmann:2009ne}
F.~Bruckmann,
\newblock PoS {\bf CONFINEMENT8}, 179 (2008), 0901.0987.

\bibitem{Polyakov:1978vu}
A.~M. Polyakov,
\newblock Phys. Lett. {\bf B72}, 477 (1978).

\bibitem{Digal:2003jc}
S.~Digal, S.~Fortunato, and P.~Petreczky,
\newblock Phys.Rev. {\bf D68}, 034008 (2003), hep-lat/0304017.

\bibitem{Lucini:2005vg}
B.~Lucini, M.~Teper, and U.~Wenger,
\newblock JHEP {\bf 02}, 033 (2005), hep-lat/0502003.

\bibitem{Lee:2009}
H.~Lee and W.~Cai,
\newblock  {\bf 2009},
\newblock Lecture Notes, Stanford University, 2009.

\bibitem{Essmann:1995}
U.~Essmann {\em et~al.},
\newblock J. Chem. Phys. {\bf 103}, 8577 (1995).

\bibitem{Lenz:2003jp}
F.~Lenz, J.~W. Negele, and M.~Thies,
\newblock Phys. Rev. {\bf D69}, 074009 (2004), hep-th/0306105.

\bibitem{Lenz:2007st}
F.~Lenz, J.~W. Negele, and M.~Thies,
\newblock Annals Phys. {\bf 323}, 1536 (2008), 0708.1687.

\bibitem{Wagner:2006qn}
M.~Wagner,
\newblock Phys. Rev. {\bf D75}, 016004 (2007), hep-ph/0608090.

\bibitem{Szasz:2008qk}
C.~Szasz and M.~Wagner,
\newblock Phys. Rev. {\bf D78}, 036006 (2008), 0806.1977.

\end{thebibliography}

\end{document}